\documentclass[12pt]{emulateapj}
\usepackage{ulem}
\usepackage{hyperref}
\usepackage{amsmath}
\usepackage{color}

\renewcommand{\emph}{\textit}
\newcommand{\blank}[1]{}

\def\gsim{\ \raise 3pt \hbox{$\rangle$} \kern -8.5pt \raise -2pt \hbox{$\sim$}\ }
\def\lsim{\ \raise 3pt \hbox{$\langle$} \kern -8.5pt \raise -2pt \hbox{$\sim$}\ }
\def\gx{\emph{GX Simulator}}
\def\gs{\emph{GS Simulator }}

\shorttitle{GX Simulator}

\begin{document}
\title{3D Radio and X-Ray Modeling and Data Analysis Software: Revealing Flare Complexity }

\author{Gelu M. Nita\altaffilmark{1}, Gregory D. Fleishman\altaffilmark{1,2}, Alexey A. Kuznetsov\altaffilmark{3}, Eduard P. Kontar\altaffilmark{4}, and Dale E. Gary\altaffilmark{1}}

\altaffiltext{1}{Center For Solar-Terrestrial Research, New Jersey
Institute of Technology, Newark, NJ 07102}
\altaffiltext{2}{Ioffe  Institute, St. Petersburg 194021, Russia}
\altaffiltext{3}{Institute of Solar-Terrestrial Physics, Irkutsk 664033, Russia}
\altaffiltext{4}{School of Physics \& Astronomy, The
University of Glasgow, G12 8QQ, Scotland, United Kingdom}

\begin{abstract}
  Many problems in solar physics require analysis of imaging data obtained in multiple wavelength domains with differing spatial resolution, in a framework supplied by advanced 3D physical models. To facilitate this goal, we have undertaken a major enhancement of our IDL-based simulation tools developed earlier for modeling microwave and X-ray emission. The enhanced software architecture allows the user to (i) import photospheric magnetic field maps and perform magnetic field extrapolations to generate 3D magnetic field models, (ii) investigate the magnetic topology by interactively creating field lines and associated fluxtubes, (iii) populate the fluxtubes with user-defined nonuniform thermal plasma and anisotropic, nonuniform, nonthermal electron distributions; (iv) investigate the spatial and spectral properties of radio and X-ray emission calculated from the model, and (v) compare the model-derived images and spectra with observational data. The package integrates shared-object libraries containing fast gyrosynchrotron emission codes, IDL-based soft and hard X-ray codes, and potential and a linear force free field extrapolation routines. The package accepts user-defined radiation and magnetic field extrapolation plug-ins. We use this tool to analyze a relatively simple single-loop flare and use the model to constrain the magnetic 3D structure and spatial distribution of the fast electrons inside this loop.  We iteratively compute multi-frequency microwave and multi-energy X-ray images from realistic magnetic fluxtubes obtained from pre-flare extrapolations, and compare them with imaging data obtained by SDO, NoRH, and RHESSI. We use this event to illustrate use of the tool for general interpretation of solar flares to address disparate problems in solar physics.
\end{abstract}

\keywords{Sun: flares --- Sun: radio radiation}

\section{Introduction}
\label{intro}

Recent years have brought tremendous progress in solar observations. A fleet of space missions observes the Sun from different positions and at various spectral ranges---from long radio waves through optical and EUV to hard X-rays and gamma-rays, while numerous ground-based observatories provide daily records in the decimeter, microwave, and submillimeter radio ranges and in the optical. Although various measurements provide imaging and spectroscopy information related to different layers---photospheric, chromospheric, and various coronal levels---the observations provide only line-of-sight integrated 2D images and spectra, and the most common method of data analysis is through image comparison, i.e., analysis in a 2D domain. The 3D structure cannot be unambiguously revealed through observations alone, without considering physical models.

To remedy the situation we put forward an essentially new approach of observation-based 3D modeling, to create a realistic 3D model structure from which to calculate emission at various wavelengths, which can be adjusted to match the corresponding observed emission. In the context of solar flares, a key element of such modeling is the 3D coronal magnetic field, which currently is not accessible to direct measurement with the necessary precision. Until such measurements are available, we must rely on the photospheric/chromospheric measurements of the surface magnetic fields and fluid motions, from which various magnetic field models (e.g. potential, linear, or nonlinear force-free field extrapolations, or MHD models) are obtained. Having such a 3D magnetic data cube, one can fill a subregion (a magnetic loop, or a collection of loops) with prescribed plasma and nonthermal particle populations  and predict escaping emission in any spectral range of interest.

Although this avenue is easy to outline, it is exceedingly complicated and time-consuming
to implement in practice and, for those few events where it has been attempted in the literature \citep{Preka_Alis_1992, Kucera_etal_1993, Lee_Gary_Zirin_1994, Bastian_etal_1998, Simoes_Costa_2006,
Tzatzakis_etal_2008,  Fl_etal_2009, Simoes_Costa_2010, Fl_etal_2013, Gary_etal_2013, Costa_etal_2013, Kuznetsov_Kontar_2014}, a wide variety of methods have been used with varying level of sophistication.
To make such 3D modeling easier and more uniform, we have developed a flexible, easy-to-use modeling framework
in a form of an IDL-based widget tool, which we call \textit{GX Simulator}, based on the latest, most sophisticated, and generally applicable codes.  As more sophisticated codes become available, they can be easily added.  The software is freely available for installation
via the solar software (SSW) distribution website\footnote{Solar software \url{www.lmsal.com/solarsoft/}}. A detailed Help file is provided in the installation package, and a web version of it is accessible online with no need for prior installation of the package.\footnote{GX Simulator online documentation \\ \url{http://web.njit.edu/ ~gnita/gx_simulator_help/}}.

In this paper, we outline the rationale, structure, and method of implementation of \textit{GX Simulator}, as well as its main features and use. Further, we demonstrate its use and functionality for a well-observed solar flare. Our analysis of this solar flare concentrates on the 3D structure of the flaring loops, their lengths and heights determined from the modeling, 3D spatial relationships, and properties of the accelerated electron distribution (energy spectrum and 3D spatial distribution) needed to consistently interpret the available X-ray and radio data.

\section{\textit{GX Simulator} overview}

\begin{figure}
\begin{center}
\includegraphics[width=0.79\columnwidth,angle=0]{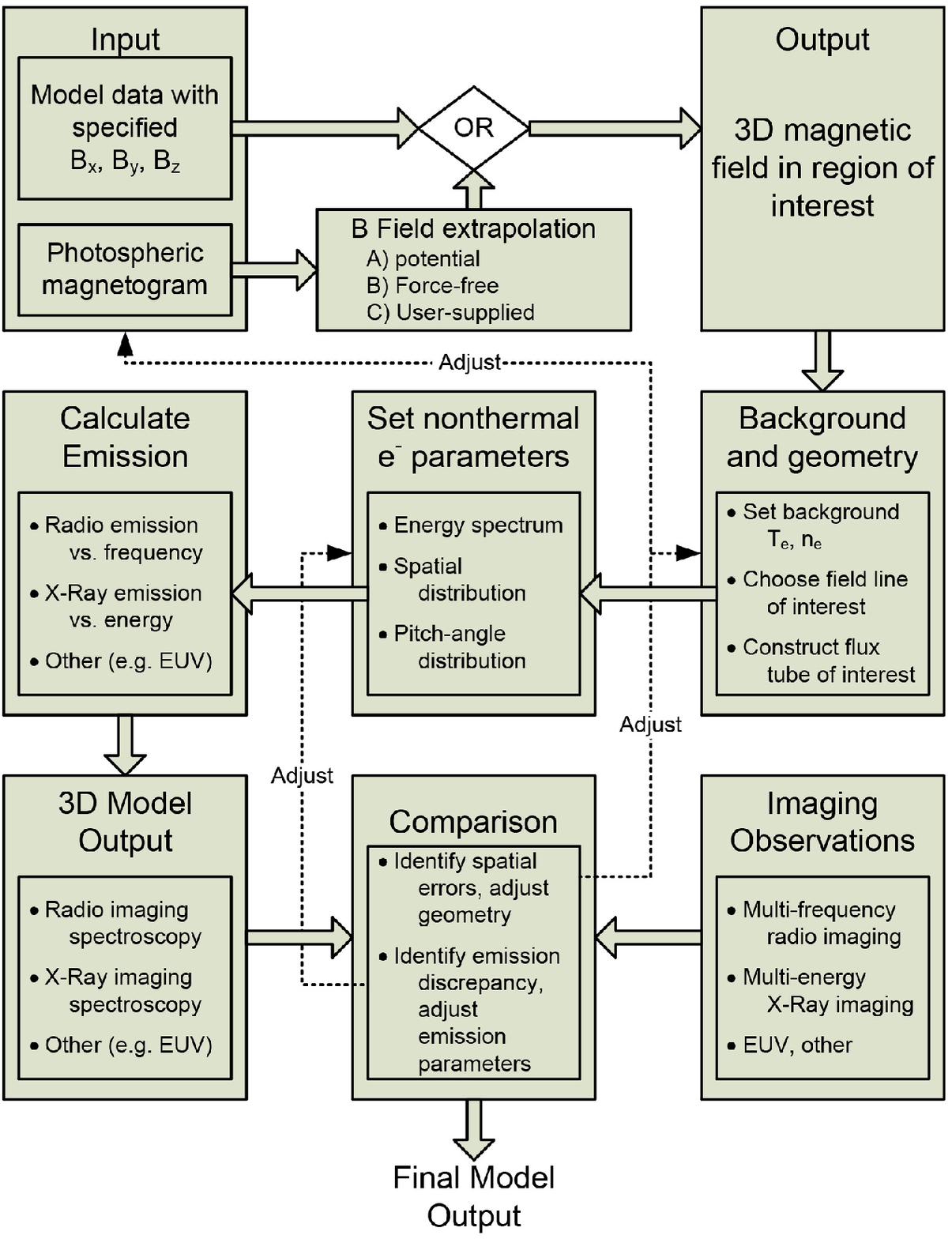}
\end{center}
\caption{\label{fig:chart}  Overview of the workflow in \emph{GX Simulator}. }
\end{figure}

As noted in the introduction, \emph{GX Simulator} is designed to simplify the modeling of 3D magnetic field, plasma and energetic particle distributions in the solar atmosphere for use in comparative analysis of 2D spatial plus spectral observational data. The overall workflow of \emph{GX Simulator} is presented in Figure \ref{fig:chart}.
The starting point of \emph{GX Simulator} is the 3D magnetic field model in the corona (see section \ref{sec:mag}),
which can be either loaded directly or modeled/extrapolated within \emph{GX Simulator}. The simple
hydrostatic background solar plasma model with uniform base pressure is automatically assumed. To this background plasma the user adds a model of the non-thermal electrons in a selected sub-region, e.g. along the magnetic field line of interest as described in Section \ref{sec:tubes}. Using the prescribed properties of the magnetic field, plasma, non-thermal particles and selected field-of-view, \emph{GX Simulator} calculates radio (section \ref{sec:radio}) and X-ray (section \ref{sec:x-ray}) imaging spectroscopy datacubes. These simulated images and spectra can be compared with the observed ones, and field line selection and particle distributions can be adjusted until an acceptable match is found.

\begin{figure}
\begin{center}
\includegraphics[width=0.59\columnwidth,angle=0]{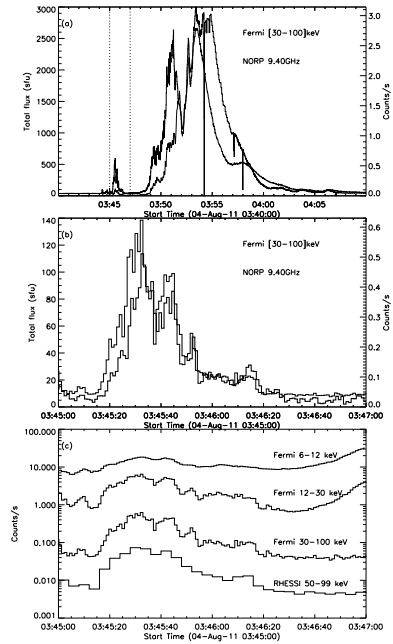}
\end{center}
\caption{\label{overview}  Overview of data available for the 4-Aug-2011 M9.3 solar flare. (a) Fermi GBM [30-100]keV (thin lines) and NoRP 9.4GHz (thick lines) lightcurves for the entire duration of the flare.
 (b) Zoom into the 2-minute time window marked by the vertical dotted lines shown in panel (a), which corresponds to the precursor of the main flare that was the only portion of this event observed by the RHESSI spacecraft. (c) Fermi GBM lightcurves for three energy channels [6-12, 12-30, 30-100]keV (1 sec time resolution), RHESSI [50-99]keV (4 sec time resolution), and Konus-\textit{Wind} lightcurve [78-312]~keV (courtesy of the Konus-\textit{Wind} team).}
\end{figure}

\begin{figure}[!th]
\begin{center}
\includegraphics[width=0.48\columnwidth,angle=0]{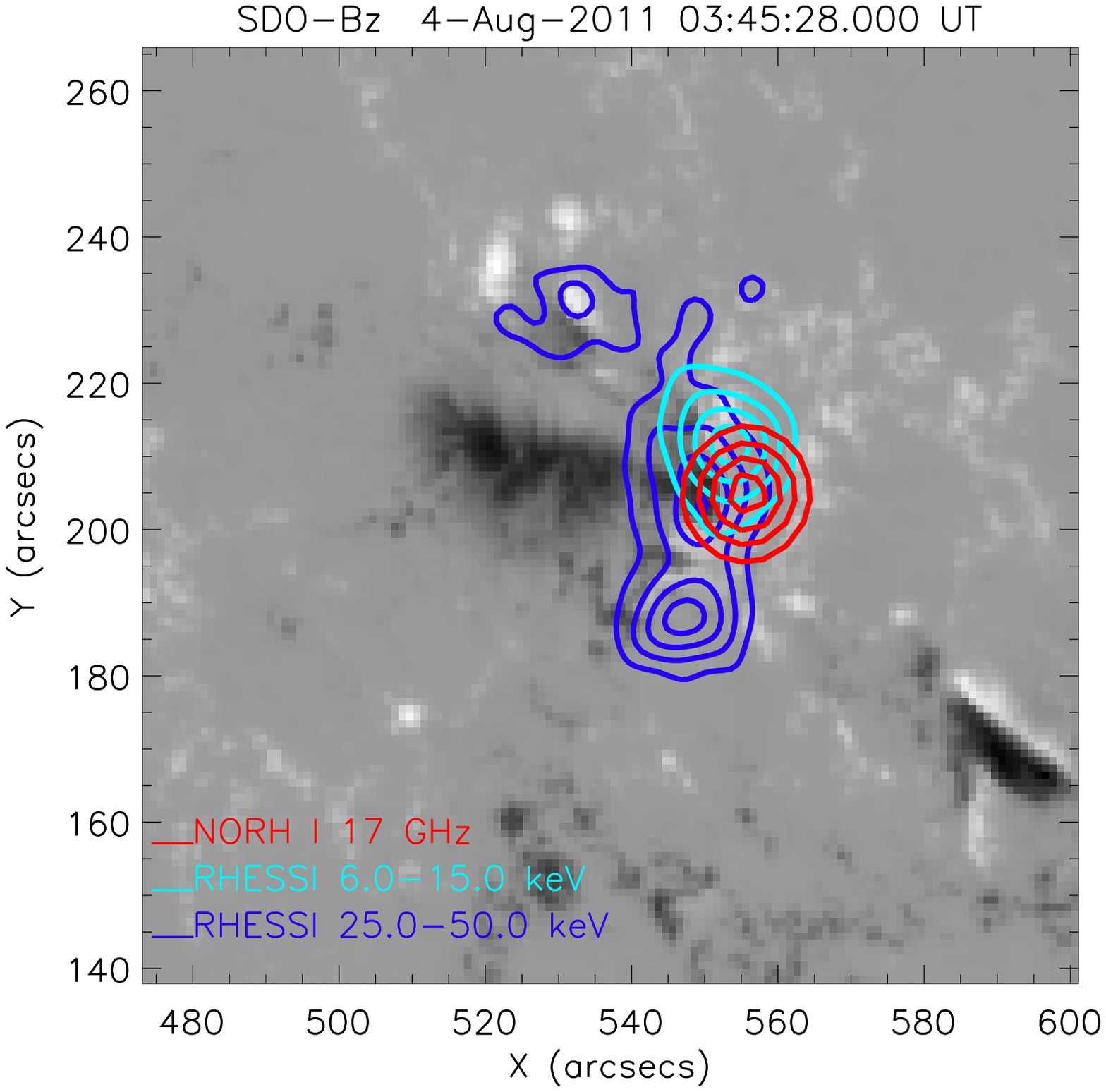}
\includegraphics[width=0.48\columnwidth,angle=0]{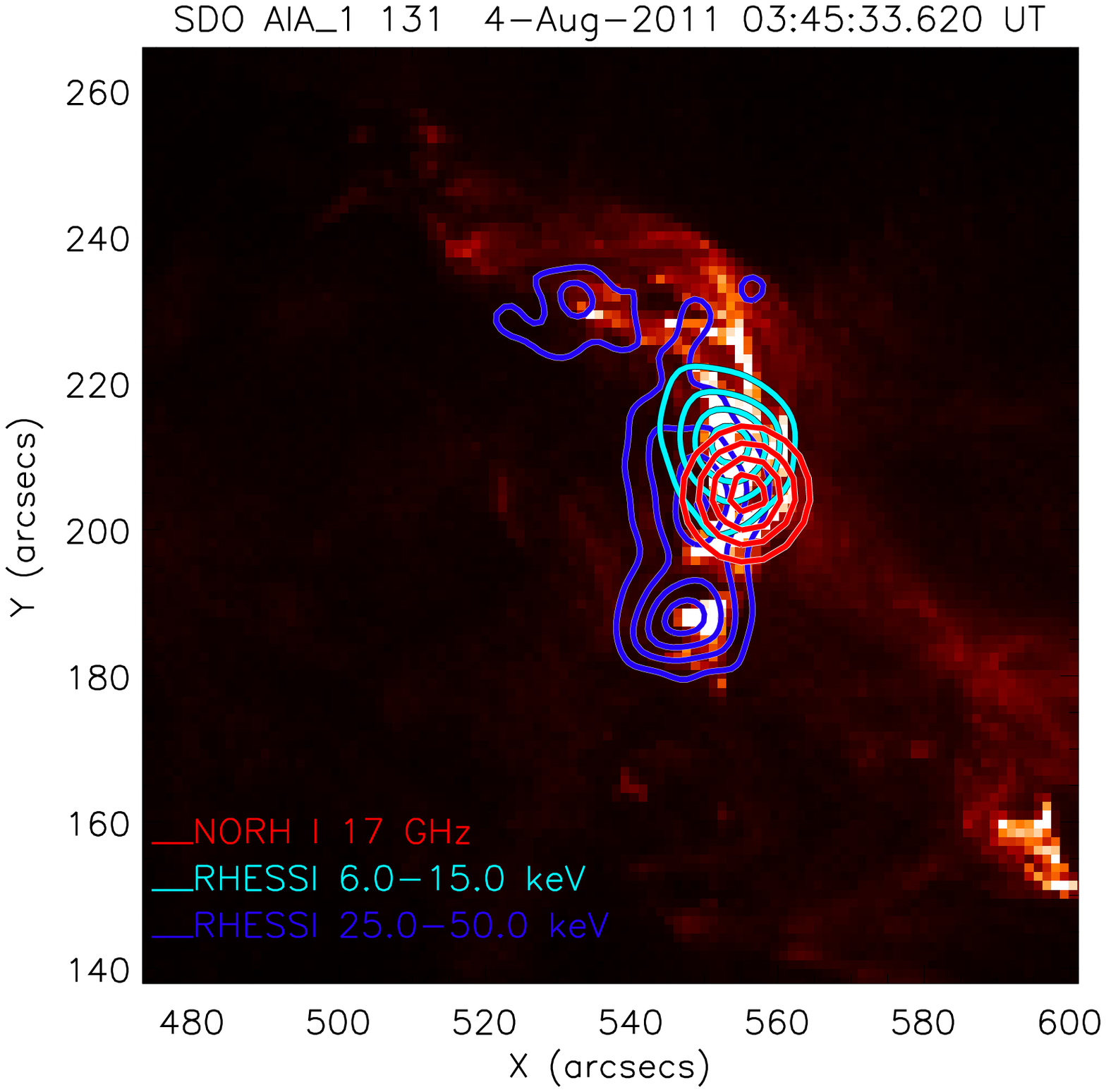}
\end{center}
\caption{\label{datamaps} Imaging observations for the 4-August-2011 event around 03:45:30 UT. Left panel: SDO HMI LOS magnetic field map. Right panel: SDO AIA 131 \AA\ EUV map. The 30, 50, 70, and 90\% contours are overlayed on each map for NoRH 17GHz Stokes I (red), RHESSI 6-15 keV (cyan) and 25-50 keV (blue). All overlayed maps are corrected for solar rotation to agree with the reference background map times shown in the plot titles.}
\end{figure}

\emph{Example:} To demonstrate the capabilities of \emph{GX Simulator}, throughout this paper we will
refer by way of example to a well observed M9.3 solar flare that occurred on 4 Aug 2011. Figures~\ref{overview}~and~\ref{datamaps} present an overview of data available for this solar flare
from RHESSI \citep[][]{2002SoPh..210....3L}, Fermi GBM \citep[][]{
2009ApJ...702..791M}, Konus-\textit{Wind}\footnote{A list of the Konus-\textit{Wind} solar flare triggers and
plots of their light curves are available at \url{http://www.ioffe.ru/LEA/Solar/.}} \citep{Palshin_etal_2014}, NoRP \citep[][]{1985PASJ...37..163N}, NoRH \citep[][]{1994IEEEP..82..705N}, and SDO\citep[][]{2012SoPh..275....3P}. Specifically, we concentrate on the first flare episode bounded by two vertical dotted lines in Figure~\ref{overview}, primarily because RHESSI missed the main flare, but also because the radio emission is dominated by a single source during this period, but grows more complex during the later period when the flux levels are exceptionally large, which would complicate the analysis.

The X-ray and radio source morphology is consistent with the standard picture of a single flaring loop producing the coronal Soft X-Ray (SXR) and radio emission and accompanying footpoint Hard X-ray (HXR) emission from
precipitating fast electrons
\citep[e.g.][as the recent reviews]{2011SSRv..159..107H,2011SSRv..159..301K,2011SSRv..159..225W}.
Inspection of the photospheric magnetogram, however, is inconclusive for identification of the possible footpoint regions: although one of the HXR sources does project onto a region of enhanced magnetic field, the southern HXR source projects onto an area without strong magnetic field. To clarify the magnetic connectivity in the event, we employed the built-in capability of the \gx \ to produce a potential (PFE) or linear force-free field extrapolation (LFFFE) from the photospheric boundary to the coronal volume
\citep[see][ for extrapolation discussions]{2005A&A...433..701W}; the extrapolated datacubes are located in the $\sim/gx\_simulator/demo/box$ SSW distribution folder.
As will be shown in more detail in the following sections, in the case of a PFE magnetic model we were unable to identify any field line subset that could have corresponded to a flare loop with footpoints at the HXR source positions and a looptop at the SXR/radio source positions. Instead, the extrapolated magnetic field lines were found to be almost transverse to the `loop' implied by the X-ray and radio morphology. However, the PFE magnetic connectivity was found to be in a good agreement with a subset of the EUV loops observed in the \citep[SDO AIA; ][]{2012SoPh..275...17L} data, Figure~\ref{datamaps} right panel, with the `radio loop' almost coinciding with one of fainter EUV loops.
Nevertheless, using an alternative LFFFE magnetic model obtained by tuning the free parameter $\alpha$ to a value of $6.8\times10^{-10}\mathrm{cm}^{-1}$, we were able to produce a magnetic topology more consistent
with the strong EUV emission observed in the SDO AIA background map, which appear
to be more or less collocated with the neutral line of the active region.

\section{Magnetic Field Models}\label{sec:mag}

\subsection{Importing Magnetic Field Models}
\gx\  may import externally produced magnetic field models from a binary IDL \emph{sav} file containing an arbitrarily named IDL structure that has a minimum set of mandatory and other optional
tags.\footnote{\url{http://web.njit.edu/~gnita/} \\ \url{gx_simulator_help/scr/Magnetic\%20Field\%20Models.htm}}
The mandatory tags are either a set of three dimensional arrays, $B_x$, $B_y$, and $B_z$, holding the vector components ($B_x$, $B_y$, $B_z$) of the magnetic field in each volume element (voxel) of the 3D model, or a four dimensional $B$ array, where the fourth dimension indexes the cartesian components of the magnetic field. As detailed in the accompanying help file distributed with the \gx\ installation package, other optional but highly recommended tags may include information related to the spatial resolution of the magnetic model (if different from the standard $2\arcsec$ MDI resolution), location of the base map on the solar disk, as well as a set of reference IDL map structures or objects corresponding to the base field of view of the model. If at least one reference map is provided, and no specific date and location tags are provided, the date, time and location of the model
are inherited from the corresponding tags present in the map structure.

Alternatively, as detailed below, one may create a magnetic field model from extrapolation of photospheric magnetic field measurements. For the example solar flare event, we used an SDO line-of-sight magnetogram obtained
for 4-Aug-2011, 03:48:00~UT.

\subsection{Creating Magnetic Field Models}
The \gx\  provides built-in capabilities for generating magnetic field extrapolation models based on input photospheric magnetic field maps imported from standard fits files or IDL \emph{save} files that contain SolarSoft (\emph{SSW}) map structures or objects. Currently the \gx\ distribution package includes\footnote{Field extrapolation: \url{http://web.njit.edu/~gnita/} \\ \url{gx_simulator_help/scr/Creating_Magnetic_Field_Models.htm}}
\begin{itemize}
\item \emph{Potential field extrapolation}: this routine calls an external DLL based on an original FORTRAN code \citep{Abramenko_1986} developed by V. Abramenko and V. Yurchishin

\item \emph{Linear Force-Free Field extrapolation}: extrapolation routine \citep{Costa_etal_2005} developed in IDL by J. E. R. Costa and T. S. N. Pinto.

\item \emph{Standardized calling procedure for user-provided field extrapolation}: provides hooks for use of alternative extrapolation methods.
\end{itemize}
The flexible IDL external-routine-calling protocol allows the user to replace the built-in extrapolation engines with any IDL-based or external code that outputs a magnetic datacube derived from an input LOS magnetogram or a three-dimensional array of map structures containing the input vector magnetogram.

The Extrapolation Project section of the \gx\  also allows the user to import any number of additional reference maps that can be rotated and re-formatted to match the time of interest and the user-defined field of view and spatial resolution.

\begin{figure}
\begin{center}
\includegraphics[width=0.99\columnwidth,angle=0]{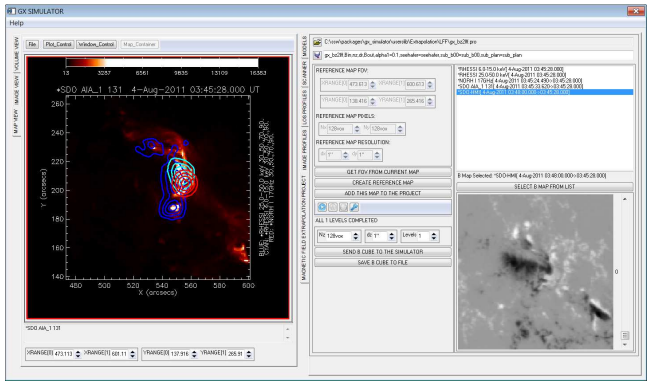}
\end{center}
\caption{\label{extrapolation} Magnetic Field Extrapolation Project view of the \gx. Left section: Plotman interactive data display showing the superposition of the same map sets displayed in Figure \ref{datamaps}b. Right Section: Collection of tools allowing the user to: a) select the field of view and resolution of the reference maps. b) add data maps to the reference map list. c) flag the photospheric magnetic field map to be used for extrapolation. d) select the IDL wrapper routine calling the extrapolation code. e) edit the default header parameters pertaining to extrapolation routine. f) select the height and resolution of the extrapolated magnetic field. g) run, pause, cancel, the extrapolation code in a parallel thread, or in debug mode. h) inspect the individual 2D layers of the performed extrapolation. i) send the magnetic model to the 3D-visualization section of the \gx.}
\end{figure}

\emph{Example:} Figure \ref{extrapolation} illustrates the use of the \gx\  Magnetic Field Extrapolation Project (MFEP) subsystem in the case of the 4 Aug 2011 flare. The left page of the MFEP user interface incorporates a customized \emph{SSW} Plot Manager (\emph{Plotman}) interactive display that allows flexible manipulation of the imported images, which may be optionally included in the list of reference maps listed on the right-hand panel. The field of view and spatial resolution of the reference maps are set using graphical controls or numerical input fields. Including reference maps in the project is optional, but the reference list must contain at least one photospheric LOS or vector magnetic field map as input to the extrapolation routine. In our case, we have imported images from SDO AIA, NoRH, and two RHESSI energies as reference maps, and used a LOS SDO HMI magnetogram taken on 4 Aug 2011 at 03:48:00~UT as the extrapolation source, rotated to the time of the RHESSI images produced for 03:45:28~UT. The IDL wrapper routine of the desired extrapolation engine is selected using a file upload control. This action exposes in a text field the header of the selected routine, which may be directly edited before its run-time compilation. The magnetic data cube output is displayed as a set of selectable, stacked 2D images on the bottom-right area of the project view. Once the MFEP output is accepted, the magnetic datacube model can be sent to the 3D-visualization subsystem of the \gx, where the magnetic field topology may be inspected, and other properties or objects may be interactively added to the model, as described in the following section.

\subsection{Inspection of the Magnetic Field Model Topology and Creation of Magnetic Flux Tubes}
The next stage in preparing the model is to use \gx\ tools to inspect the magnetic field topology and select a subset of magnetic field lines to serve as the locus of flare-accelerated particles.  If no suitable magnetic field lines seem to match the topology implied by observations, it may be necessary to go back to the previous step to adjust the magnetic field extrapolation within allowed constraints.

\begin{figure}
\begin{center}
\includegraphics[width=0.4\columnwidth,angle=0]{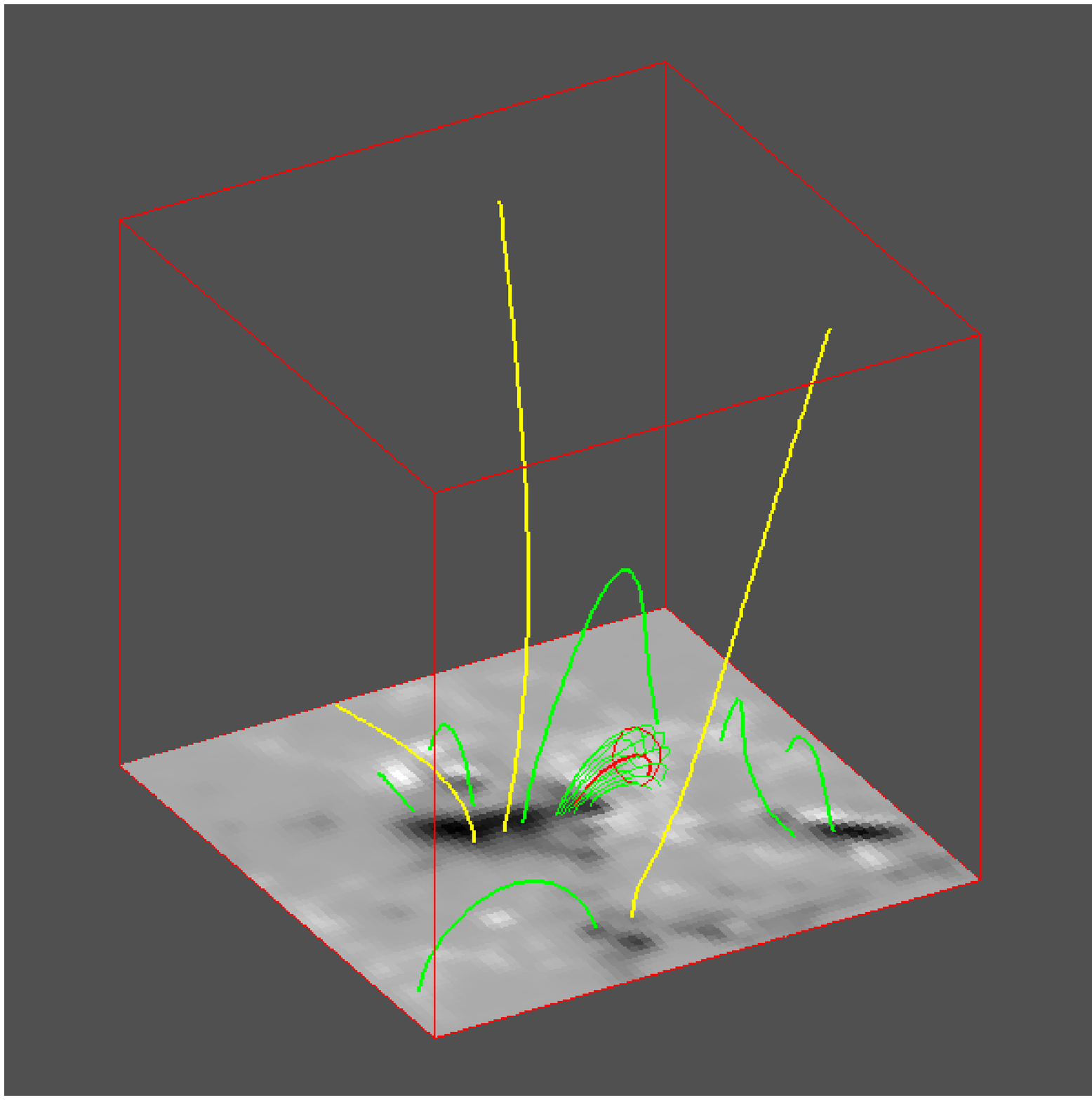}
\includegraphics[width=0.4\columnwidth,angle=0]{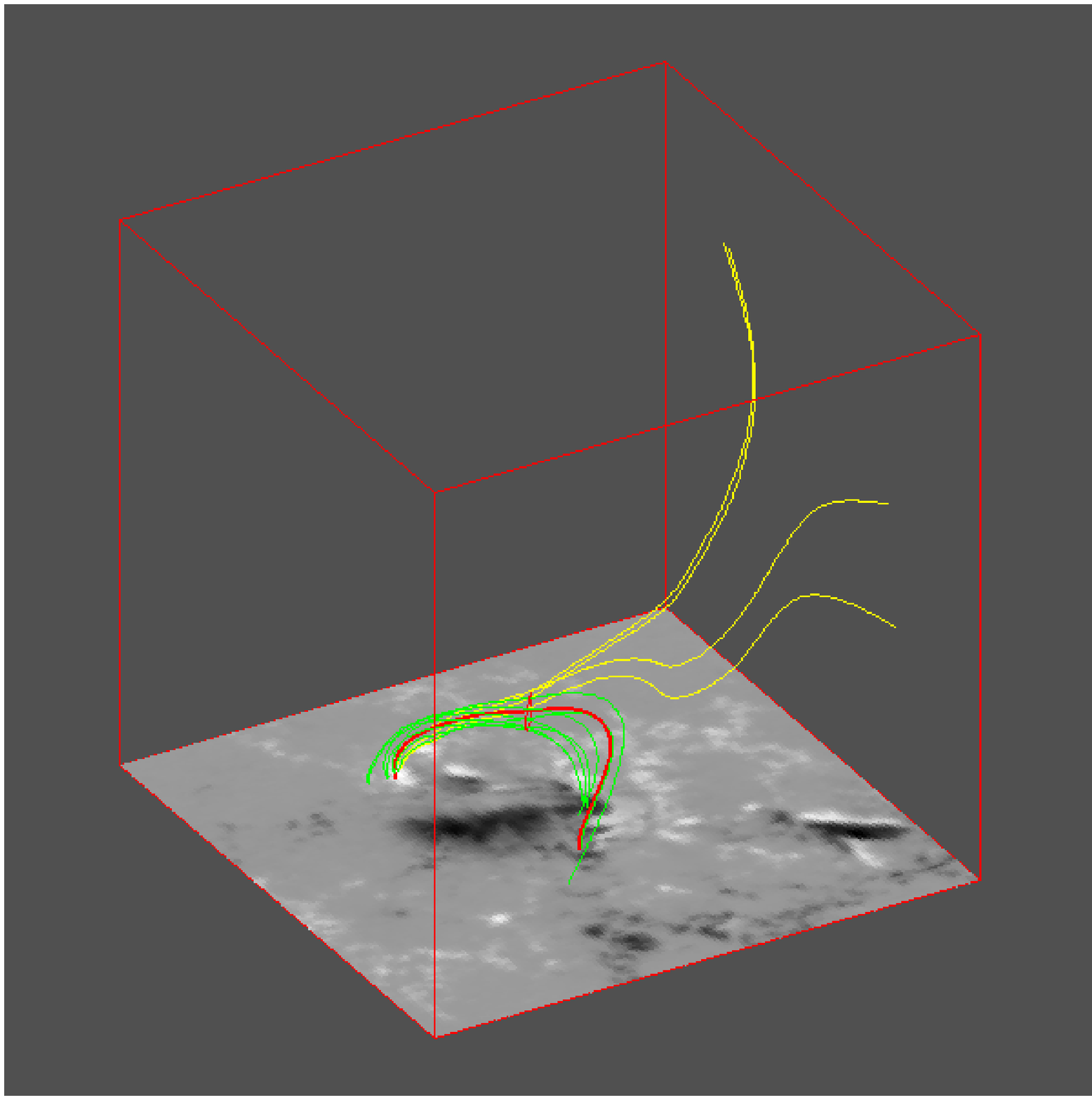}\\
\includegraphics[width=0.4\columnwidth,angle=0]{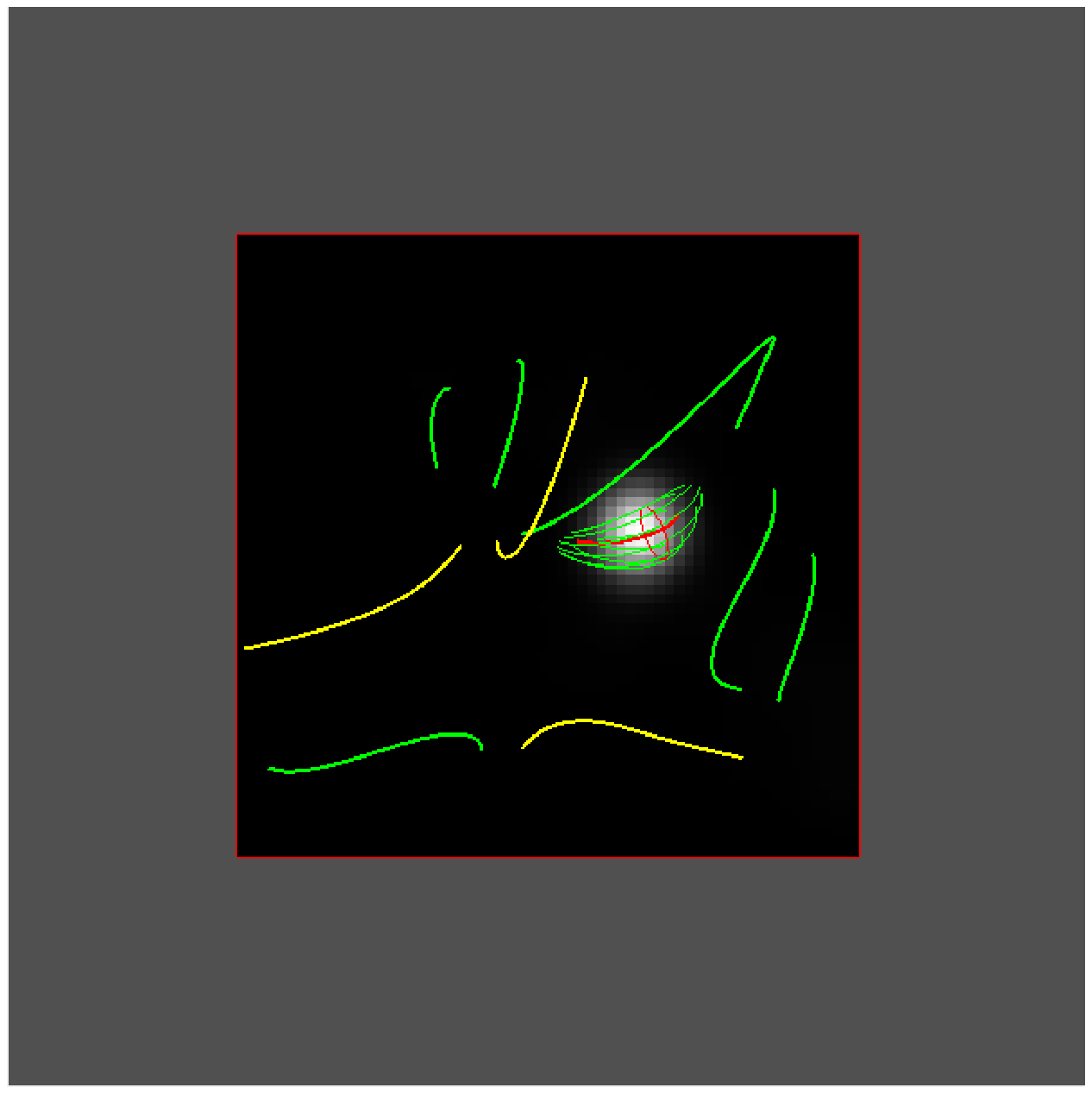}
\includegraphics[width=0.4\columnwidth,angle=0]{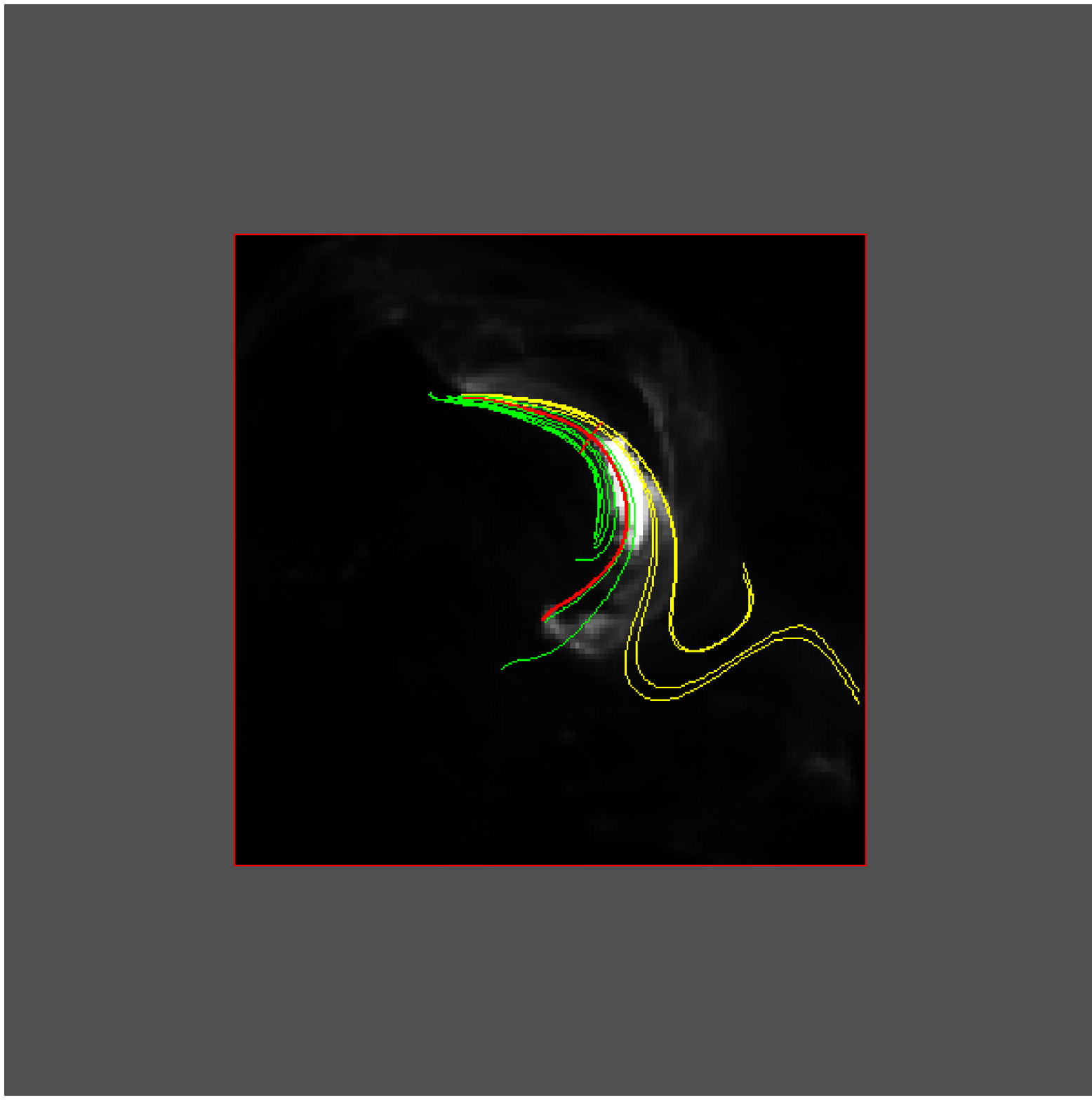}\\
\includegraphics[width=0.4\columnwidth,angle=0]{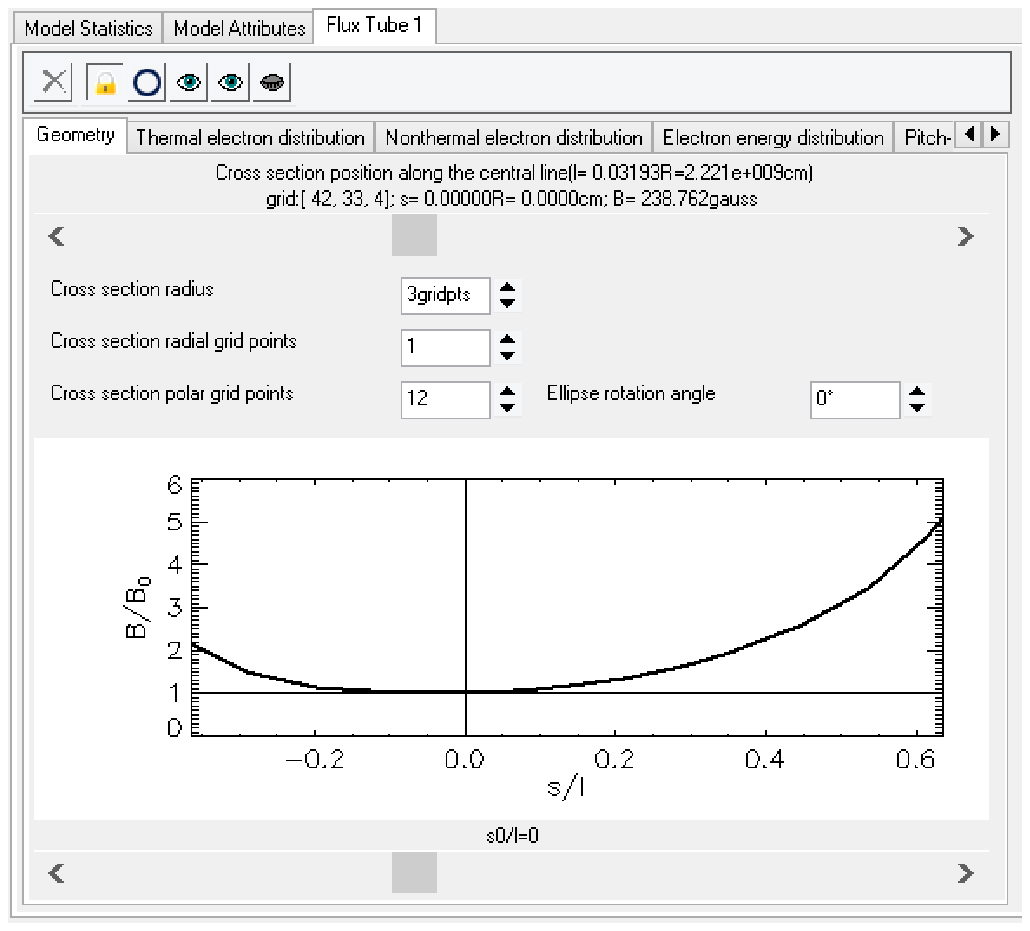}
\includegraphics[width=0.4\columnwidth,angle=0]{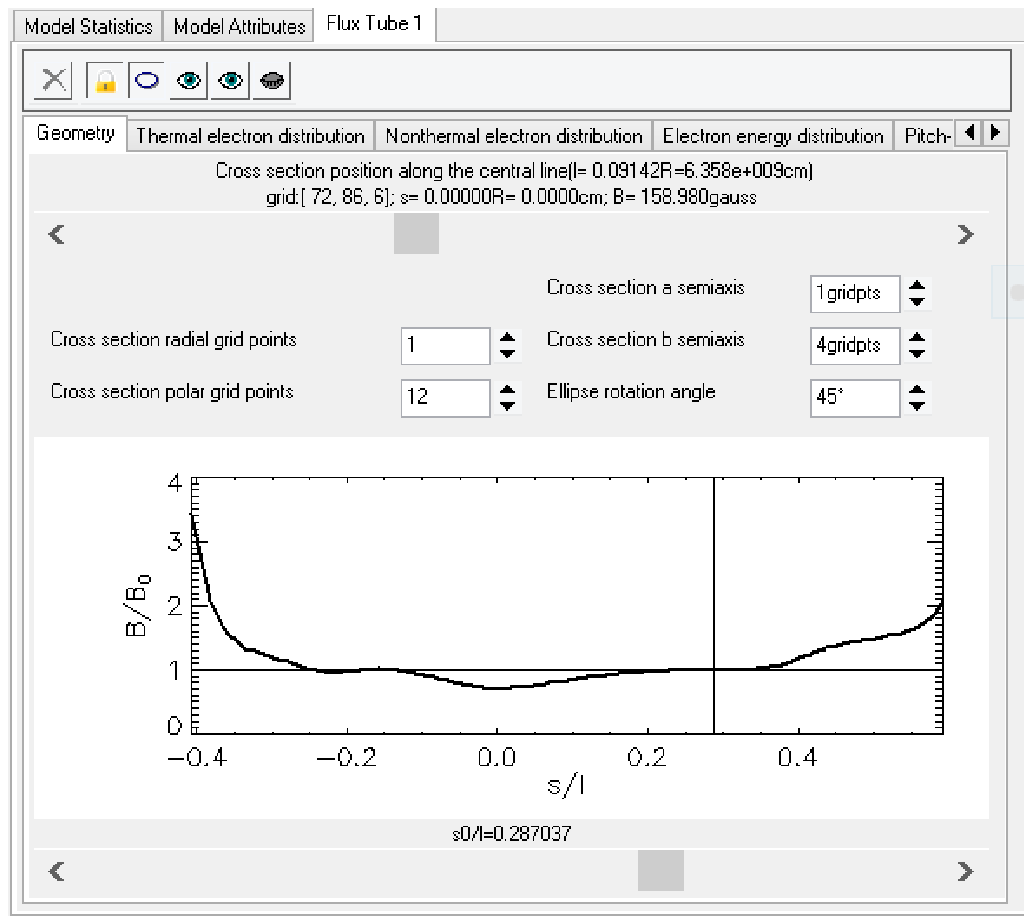}
\end{center}
\caption{\label{fieldline_view}
Magnetic field topology inspection and fluxtube selection in the case of  PFE (left column) and LFFFE (right column). Top row: Closed field line (green) and open field line (yellow) perspective view with photospheric $B_z$ magnetic field map at the bottom of the extrapolated magnetic cube. Middle row: Top view against the NoRH 17 GHz radio map (left panel) and SDO AIA\blank{\_1} 131 map (right panel). Bottom row: settings to control the fluxtube geometry.}
\end{figure}

Figure \ref{fieldline_view} illustrates the 3D visualization and manipulation capabilities of \gx\ in the case of the PF extrapolation (left column) and LFFF extrapolation (right column). The top row of Figure \ref{fieldline_view} shows a set of field lines corresponding to the PFE (left panel) and LFFFE (right panel) magnetic models from a 3D perspective. \gx\  allows the user to interactively create such field lines by mouse selection (clicking on locations on the photospheric magnetic field maps), or by numerical specification of any ($x,y,z$) coordinate within the 3D volume. The magnetic field line passing through the selected point is then drawn on the display, colored yellow if it leaves the volume through the side or top boundaries, and green if it closes within the volume at the bottom boundary.  The middle row of Figure~\ref{fieldline_view} shows the same field lines from a top view\footnote{When performing data to model comparison, it is essential to take into consideration the fact that the reference data emission maps, as well as the base magnetic field map used by \gx~to generate a PFE or LFFE magnetic cube model, are LOS images, even if the center of the rectangular field of view is not located at the disc center. Thus, forcing a top view perspective when generating the synthetic maps is needed in order to obtain the correct alignment of these maps with the perspective from which data were obtained.} perspective against one of the available reference maps, a radio map for the PFE model, and an EUV map for the LFFFE model. From such a display, it is possible to visually identify one or more suitable field lines as the ``spine'' of a flaring loop.  Using the \gx\ tools one can then construct a flux tube structure as illustrated by the red central field line and adjacent green field lines in Figure~\ref{fieldline_view}. The bottom panels display the interactive input fields that \gx\  dynamically creates in order to control the geometrical properties of a flux tube: the position $s$ along the central field line (sliding control) of a reference cross-section surface and its elliptical (left panel) or circular (right panel) cross-section defined in model grid units. A second slider control (bottom of the page) allows the user to select the longitudinal coordinate $s_0$ of the reference normal cross section from which all non-central flux tube field lines (green color) originate. When a new flux tube is created, the reference cross section coincides by default with the loop apex. The geometry control panel also displays information about the central field line characteristics such as length, loop-top position and magnetic field strength. The plot of the magnetic field strength relative to the reference values corresponding to the user defined coordinate $s_0$ is also displayed.

\emph{Example:} In the case of the 4 Aug 2011 flare, we are able to reject the PF extrapolation result and refine the range of acceptable $\alpha$ in the LFFF extrapolation based on comparison with observations. Although Figure~\ref{datamaps} suggests a simple topology, Figure \ref{fieldline_view} reveals that the PFE magnetic field lines at the flare location are almost transverse to the `loop' suggested by the combined X-ray, radio, and EUV morphology (although we note that the PFE magnetic connectivity was found to be in good agreement with other EUV loops observed elsewhere in the active region). At the same time, the alternative LFFFE magnetic model corresponding to $\alpha=6.8\times10^{-10}\mathrm{cm}^{-1}$ provides a more sheared magnetic topology consistent with the shape of the strong EUV emission observed in the SDO background map, which appears to follow the neutral line of the active region.  The value of $\alpha$ was arrived at by trial-and-error comparison with the observed topology and is not claimed to be unique, but it is illustrative of the relative ease with which \gx\ permits the use of observational data to constrain magnetic extrapolations.

\section{Magnetic flux tube models: thermal plasma and non-thermal electrons }\label{sec:tubes}

For each user-defined magnetic flux tube, \gx\  dynamically creates a new \emph{Flux Tube} control panel containing a series of tabbed pages that allow the user to interactively control the main physical properties of the corresponding magnetic flux tube model. These properties include, but are not limited to, thermal and nonthermal particle density distributions as well as distributions over energy and pitch-angle, as illustrated in the following sub-sections.  These distributions are enhancements added to the background hydrostatic atmosphere mentioned earlier. The fully populated models employed in this paper are provided in the $\sim/gx\_simulator/demo/gxm$ SSW distribution folder.

\subsection{Thermal plasma}\label{subsec:thermal}

The thermal electron density distribution (e.g. that of a single-temperature ``super-hot'' component) is analytically defined as a product of a normalization factor $n_0$, a unit-amplitude radial distribution $n_r$ and unit-amplitude vertical distribution $n_z$,
\begin{equation}
\label{thermal_distribution}
n_{th}(x,y,z)=n_0 n_r(x/a,y/b) n_z[z(s)/R],
\end{equation}
where $x/a$ and $y/b$ represent the flux tube cross-section cartesian coordinates $x$ and $y$ normalized by the ellipse semi-axes $a$ and $b$ of the reference cross-section intersecting the flux tube at the longitudinal coordinate $s_0$, and $z(s)$ represents the vertical coordinate corresponding to the normal cross-section containing the point $(x,y,z)$, which intersects the central field line at the longitudinal coordinate $s$. The solar radius $R$ also enters the vertical distribution expression as a spatial-scale normalization factor.

By default, \gx\  uses the predefined generalized gaussian radial distribution
\begin{eqnarray}
\label{thermal_radial}
n_r(x,y)=\exp\left[-\left(p_0\frac{ x}{a}\right)^2-\left(p_1\frac{ y}{b}\right)^2 -\left(p_2\frac{ x}{a}\right)^4-\left(p_3\frac{ y}{b}\right)^4\right],
\end{eqnarray}
where $p_i$ ($i=0\dots3$) are adjustable dimensionless parameters; by default, $p_0=p_1=0.5$ and $p_2=p_3=0$.  The default vertical distribution is a simple hydrostatic formula adapted from \S~3.1 of \citet{2004psci.book.....A},
\begin{eqnarray}
\label{thermal_vertical}
n_z\left[z(s)/R\right]=\exp\left[-\frac{z(s)/R}{6.7576\times10^{-8}T_0}\right],
\end{eqnarray}
where $T_0$ is an adjustable constant flux tube temperature.

\begin{figure}
\begin{center}
\includegraphics[width=0.48\columnwidth,angle=0]{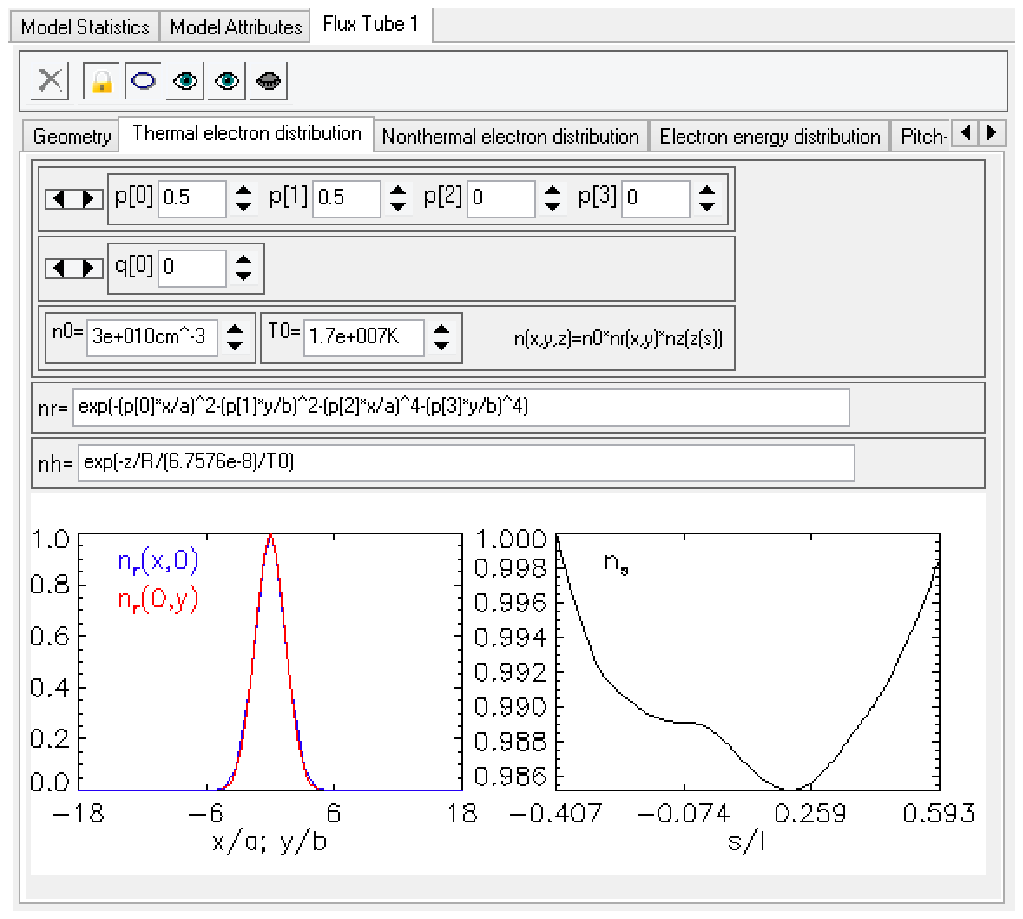}\hspace{0.05in}
\includegraphics[width=0.48\columnwidth,angle=0]{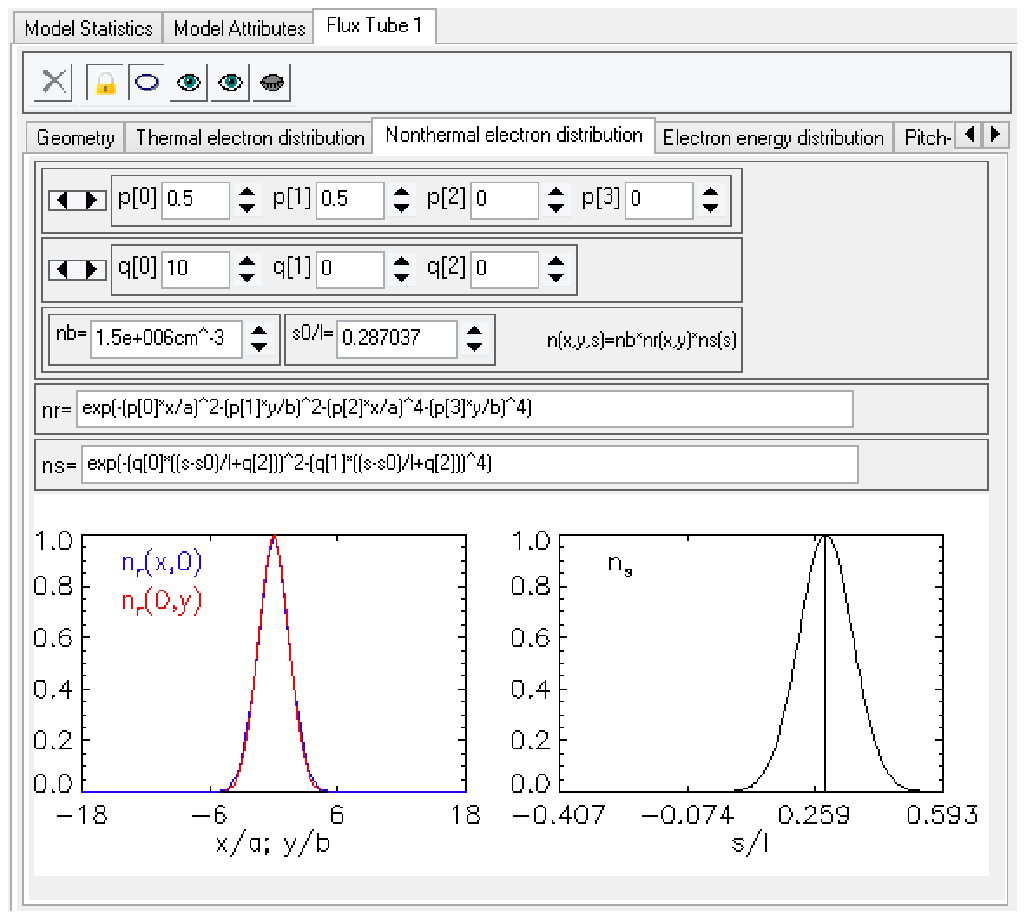}\\
\includegraphics[width=0.48\columnwidth,angle=0]{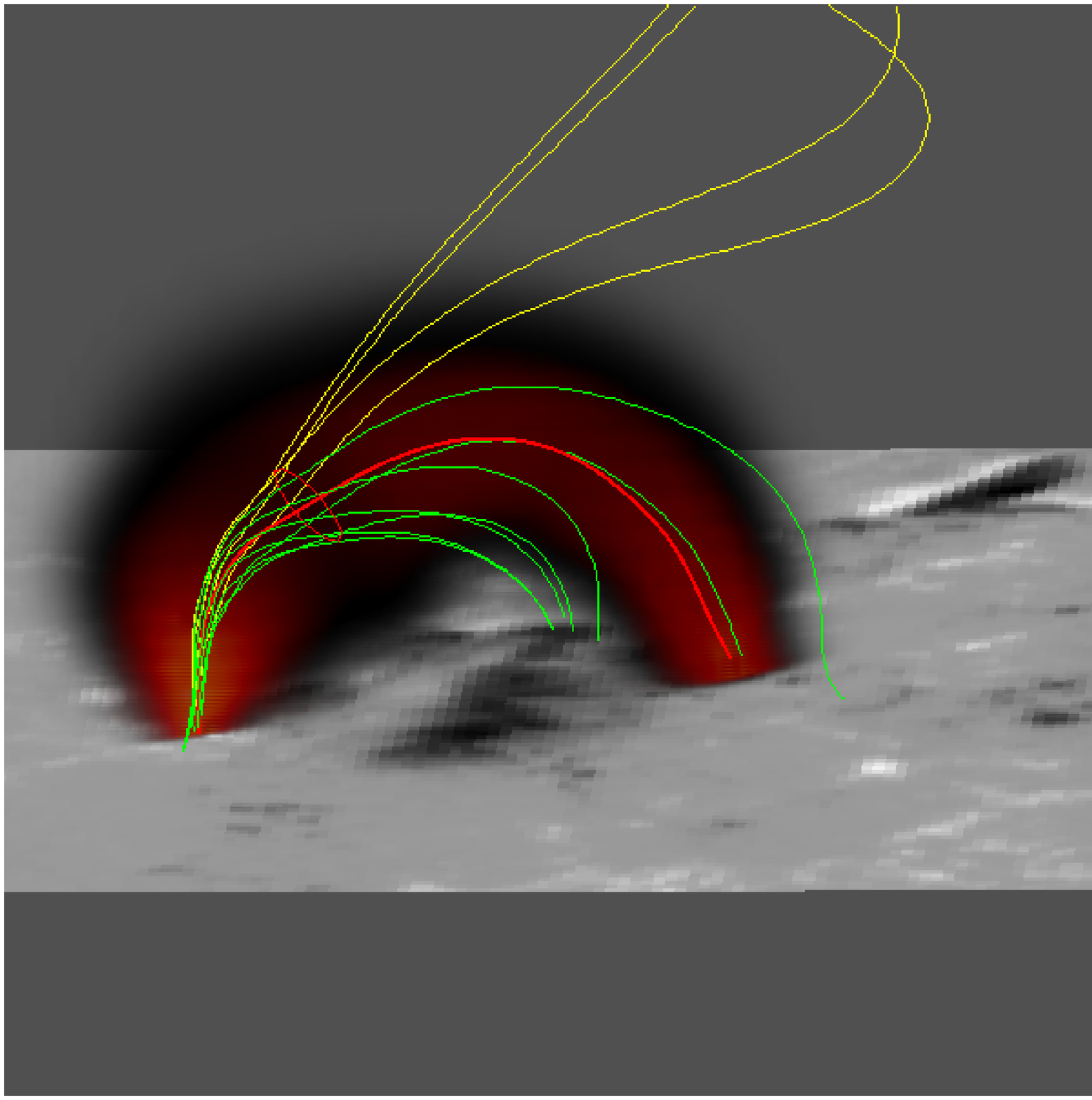}\hspace{0.05in}
\includegraphics[width=0.48\columnwidth,angle=0]{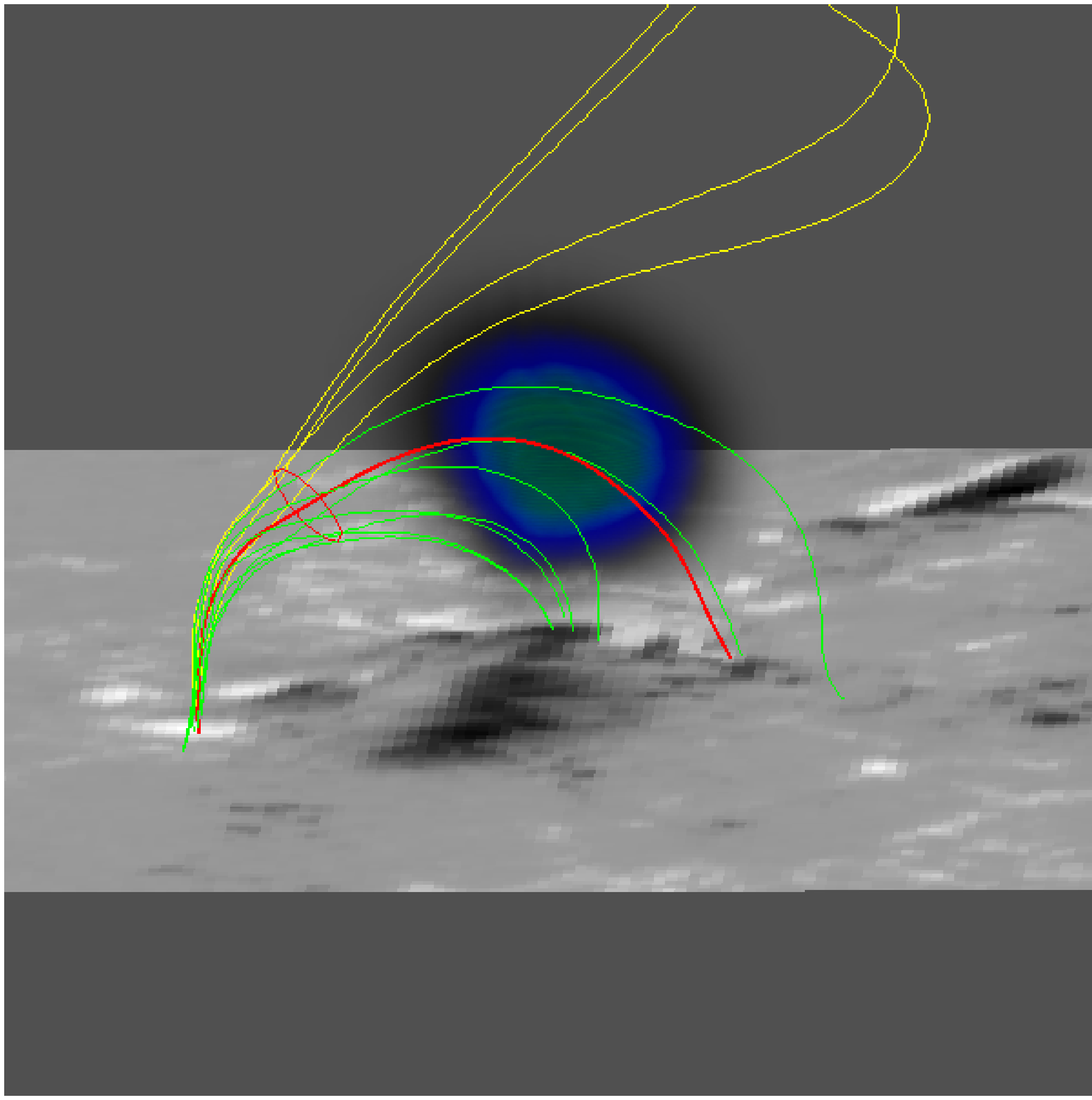}
\end{center}
\caption{\label{fluxtube_lfffe} Thermal (left panels) and non-thermal (right panels) electron spatial distributions that fill the volume of the selected LFFFE fluxtube shown in right column of Figure \ref{fieldline_view}. The definitions of numeric parameters used are given in the text.}
\end{figure}

Although the default analytical forms of these distributions were chosen to provide considerable flexibility, advanced users have the complete freedom to define their own analytical distributions. For this purpose, the functional forms of these distributions, the number of free parameters, and their values, may be edited at run-time using the textual and numerical input fields provided on their corresponding control tabs. The syntax is automatically checked for errors and a compiler message is generated if the user-defined syntax is not compliant with the IDL programming environment rules. The radial and longitudinal profiles of the user-defined thermal particle density distributions are plotted and automatically updated on the corresponding panel, as shown on the top row of Figure~\ref{fluxtube_lfffe}.

\emph{Example:} The left-hand panels of Figure~\ref{fluxtube_lfffe} present the Flux Tube panel (top) and 3D distribution (bottom) describing the thermal electron density distribution for the flux tube chosen for the LFFFE model for the 4 Aug 2011 flare.  The single temperature is about 20~MK to match the thermal component of the hard X-ray spectrum, while the distributions across the loop are chosen to roughly match the soft-X-ray source shape. One can note that within the simple hydrostatic coronal plasma model the agreement between the observed and modeled SXR source is rather poor. However, we checked that this can be improved significantly by relaxing the assumption of the hydrostatic model; we do not show this improvement as we concentrate on the nonthermal emissions at the first place.

\subsection{Nonthermal electrons}
\subsubsection{Nonthermal electrons: spatial distribution}
In analogy to equation~\ref{thermal_distribution} for the thermal electron distribution, the nonthermal electron density spatial distribution associated with a given flux tube is defined as the product of a normalization factor $n_b$, a unit-amplitude radial distribution ($n_r$) and a unit-amplitude longitudinal distribution along the central field line ($n_s$),
\begin{equation}
\label{nonthermal_distribution}
n_{nth}(x,y,z)=n_b n_r(x/a,y/b) n_s(s/l),
\end{equation}
where $a, b,$ and $s$ are defined as above, and $l$ is the length of the central field line.

The default nonthermal radial distribution used by \gx\  has the same functional form as the thermal radial distribution defined by equation~(\ref{thermal_radial}), but to allow for confinement of the nonthermal distribution near an ``injection'' site (e.g. by turbulence) the default longitudinal distribution is defined as
\begin{equation}
\label{nonthermal_along}
n_{s}(s)=\exp\left\{-\left[q_0\left(\frac{s-s_0}{l}+q_2\right)\right]^2-\left[q_1\left(\frac{s-s_0}{l}+q_2\right)\right]^4\right\},
\end{equation}
where, beside the previously defined $s$, $s_0$ and $l$ parameters, $q_0$ and $q_1$ are dimensionless scaling factors, and $q_2$ an adjustable location parameter that may be regarded as the injection point of the non-thermal particles, in case it is different from the loop apex point $s_0$.

Again, all of these default analytical forms can be changed at run-time to any desired, user-defined expressions, which are checked for syntax upon entry, updated and plotted in the panel.

\emph{Example:} The right-hand panels of Figure~\ref{fluxtube_lfffe} present the nonthermal electron distribution panel (top) and 3D distribution (bottom) describing the nonthermal electron density distribution for the flux tube chosen for the LFFFE model for the 4 Aug 2011 flare.  The rather narrow confinement of the electrons along the loop seems to be demanded by the NoRH 17GHz radio image, as well as radio and hard X-ray spectral parameters, as will be discussed later.  For completeness, we attempted to fit all observations for the 4 Aug 2011 flare with both PFE-based and LFFFE-based models, which will be contrasted in the following. Table~\ref{distrib_table} lists the actual parameters assumed in the case of PFE (left column) and the LFFFE (right column) fluxtubes; in all cases  we adopted $q_1=q_2=0$.

\begin{deluxetable}{lllll}
\tablecolumns{3}
\tablewidth{0pc}
\tablecaption{\label{distrib_table}Assumed parameters for the 4 Aug 2011 event.
}
\tablehead{\colhead{Parameter} & \colhead{PFE} & \colhead{LFFFE}}
\startdata
\cutinhead{Thermal Density Distributions}
$T_0$ & $1.8\times10^7 {\rm K}$  & $1.7\times10^7 {\rm K}$ \\
$n_0$ & $1.0\times10^{11}{\rm cm}^{-3}$  & $3.0\times10^{10}{\rm cm}^{-3}$\\
$p_0$ & 2.5 & 0.5\\
$p_1$ & 2.5 & 0.5\\
$N_0\tablenotemark{*}$ & $5.0\times10^{37}$ & $1.3\times10^{38}$\\
\cutinhead{Nonthermal Density Distributions}
$n_b$ & $1.0\times10^6{\rm cm}^{-3}$  & $1.5\times10^{6}{\rm cm}^{-3}$\\
$p_0$ & 2.5 & 0.5\\
$p_1$ & 2.5 & 0.5\\
$q_0$ & 6 & 10\\
$s_0/l$ &0 & 0.287\\
$N_b\tablenotemark{*}$ & $1.1\times10^{33}$ & $8.9\times10^{32}$
\enddata
\tablenotetext{*}{Total number of electrons corresponding to the assumed fluxtube geometry}
\end{deluxetable}

\subsubsection{Nonthermal electrons: energy distribution}

In addition to the  built-in energy distributions defined in \citet{Fl_Kuzn_2010}, and implemented in the \gs \citep{Kuznetsov_etal_2011} application, \gx\  introduces two new energy distributions, namely \emph{Thermal plus Power Law} (TPL), and \emph{Thermal plus Double Power Law} (TDL), which, as suggested by their names, consist of a thermal core and a single or double power law distribution over a limited energy range.  The power law component is specified by low and high cutoff energies and power-law-index $\delta$, as well as a break energy in the case of the double power law. The \emph{Flux Tube} user interface of the \gx\ contains an \emph{Electron Energy Distribution} control tab that allows selection of the desired distribution type and, upon selection, exposes all input parameters relevant for the selected distribution.  Note that the thermal component of TPL or TDL distributions\footnote{The difference of these two distributions compared with either single- or double power-law distributions is that the former takes the thermal electron contribution into account to compute the GS emissivity and absorption, while the latter does not.
In all other respects the choice of TPL distribution is equivalent to PWL, and the choice of TDL is equivalent to DPL. The  PWL or DPL computations are faster than, respectively, the TPL or TDL; thus, the use of PWL or DPL distributions is warranted when the contribution of the thermal plasma to the GS emissivity and opacity is small; otherwise, the TPL or TDL has to be used.} is exactly the same thermal component specified in section~\ref{subsec:thermal}.

\begin{figure}
\begin{center}
\includegraphics[width=0.48\columnwidth,angle=0]{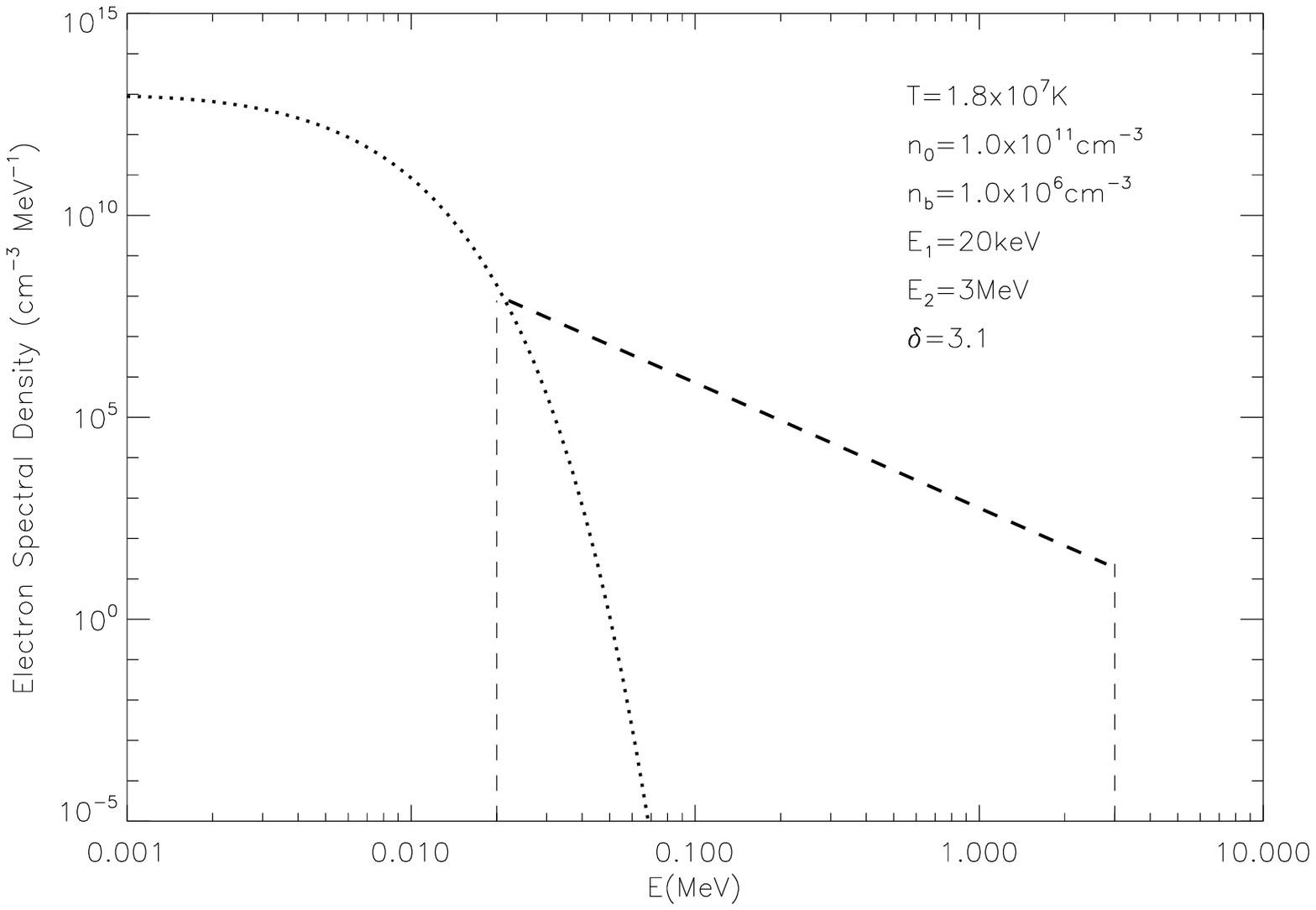}
\includegraphics[width=0.48\columnwidth,angle=0]{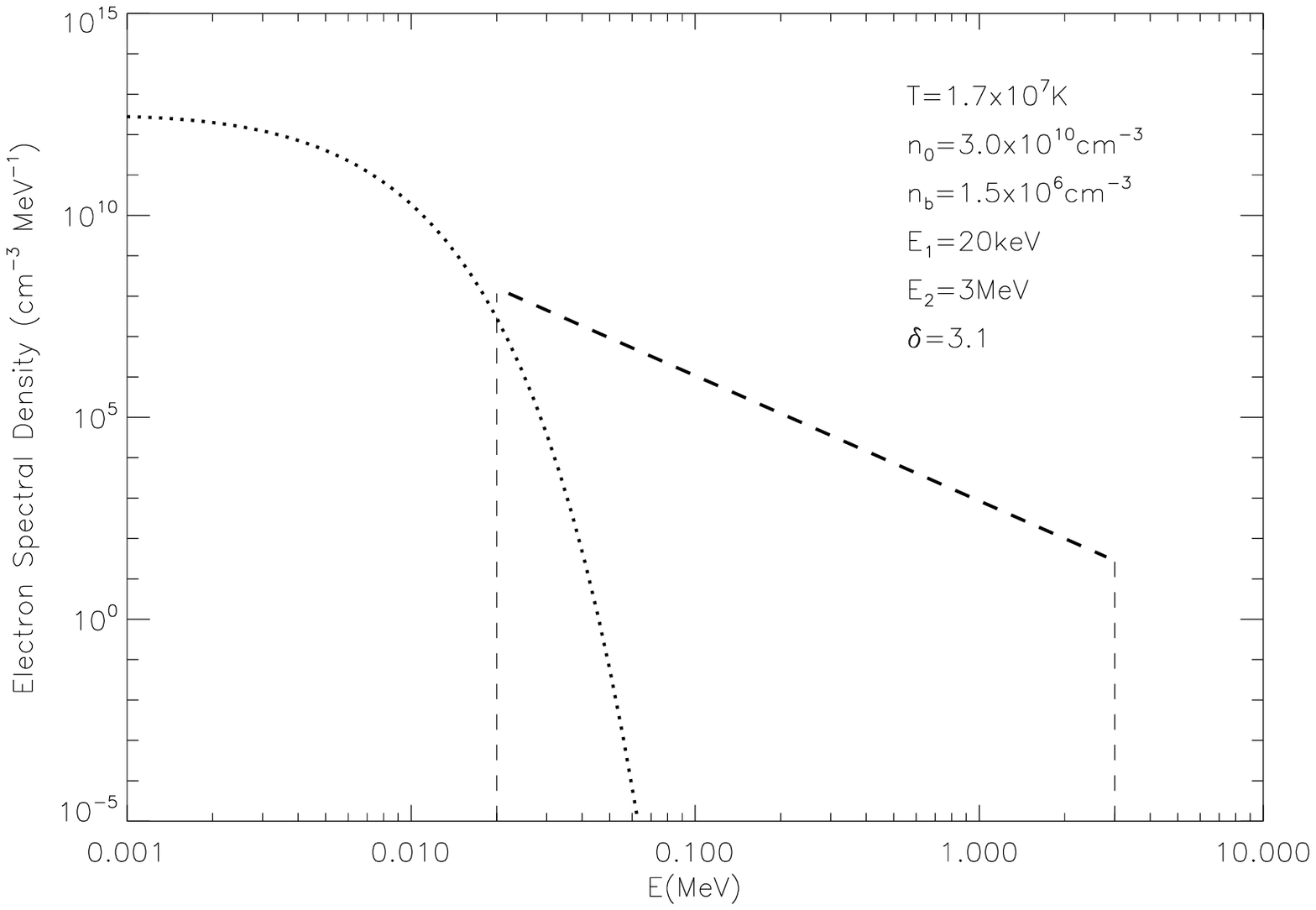}
\end{center}
\caption{\label{TPL}  Assumed TPL energy distributions for the 4 Aug 2011 flare in the case of the PFE fluxtube (left panel) and the LFFFE fluxtube (right panel). The TPL distributions were defined as the superposition of thermal distributions, characterized by the thermal electron densities $n_{th}$ and temperatures $T$, and power-law distributions over the $0.02-3$~MeV energy range, characterized by the nonthermal electron densities $n_{nth}$ and the power-law indices $\delta$. The parameters corresponding to each case are displayed on the corresponding plots.}
\end{figure}

\textit{Example:} Choice of nonthermal energy distribution for the 4 Aug 2011 flare is based on the RHESSI photon spectrum, which shows a typical thermal plus power law shape. Figure~\ref{TPL} shows the selections made for both PFE and LFFFE models, for which, as will be shown in more detail below, the TPL distribution was required to provide enough degrees of freedom to allow a satisfactory match of the radio spectra recorded by NoRP.

\subsubsection{Non-thermal electrons: pitch angle distribution}

\gx\ implements all  electron distributions over pitch angle defined in \citet{Fl_Kuzn_2010}, which include isotropic, exponential loss-cone, Gaussian loss-cone, Gaussian, and a generalized Gaussian.

\emph{Example:} For the 4 Aug 2011 event, the simplest distribution (isotropic) has been assumed in both PFE and LFFFE models because the data did not offer any clear constraint on the angular distribution.  However, if is known \citep{Fl_Meln_2003b, Fl_Meln_2003a}  that pitch angle does significantly affect radio emission directionality, and it is possible that the spatial restriction of nonthermal electrons in Figure~\ref{fluxtube_lfffe} could be relaxed by suitable choice of non-isotropic pitch angle distribution.

\section{Escaping Radiation}
Once a magnetic field model is customized and populated with plasma and non-thermal
particles, \gx\ can in principle calculate electromagnetic emission and radiation transfer for any radiation process. It currently calculates both radio and thin-target X-ray emissions. The user can choose any 3D orientation of the model for the observer LOS, as well as define the region of interest for which the escaping radiation will be computed.

\gx\  provides a flexible plug-in standard that allows any user-defined radiation transfer code to be used in place of the default ones, provided that they adhere to the calling conventions detailed in the \gx\ package help file. The radiation transfer engines use multi-dimensional LOS information provided by the general-purpose \gx\  scanning engine, which slices the 3D model along the desired LOS direction.

\subsection{Radio emission}\label{sec:radio}

\gx\  computes radio emission based on the fast gyrosynchrotron algorithm developed by \citet{Fl_Kuzn_2010}, implemented here in the  form of external libraries  optimized for the use within \gx\ and callable from IDL. This code accounts for gyrosynchrotron and free-free radio emissions and absorption within the entire magnetic data cube; outside the cube the emission propagates as in vacuum.

The radio brightness map (the model emission intensity as a function of 2D coordinates $x$ and $y$) at a given frequency $f$ is calculated by numerical integration of the radiation transfer equation, which, in addition to the volume emissivity and absorption, also includes the linear mode conversion (mode coupling) and, for a given pixel, has a form
\begin{multline}\label{rt_1}
\frac{\mathrm{d}I_{\mathrm{L, R}}(f,  z)}{\mathrm{d}z}=j_{\mathrm{L, R}}(f,  z)
-\varkappa_{\mathrm{L, R}}(f,  z) I_{\mathrm{L, R}}(f,  z)\\ -\eta_{\mathrm{L, R}}(f,  z) I_{\mathrm{L, R}}(f,  z)
+\eta_{\mathrm{R, L}}(f,  z)I_{\mathrm{ R, L}}(f,  z),
\end{multline}
along all selected lines of sight, where $I_{\mathrm{L}}$ and $I_{\mathrm{R}}$ are the spectral intensities of the left- and right- elliptically polarized emission components, respectively, $j_{\mathrm{L}}$ and $j_{\mathrm{R}}$ are the corresponding  emissivities, $\varkappa_{\mathrm{L}}$ and $\varkappa_{\mathrm{R}}$ are the absorption coefficients, and $\eta_{\mathrm{L, R}}(f,  z)$ are the factors accounting for the frequency dependent mode conversion. The term $-\eta_{\mathrm{L, R}}(f,  z) I_{\mathrm{L, R}}(f,  z)$  accounts the fraction of the given wave modes that leaks into the other wave mode, while the term $+\eta_{\mathrm{R, L}}(f,  z)I_{\mathrm{ R, L}}(f,  z)$ accounts for the fraction of the other wave mode that is converted to the given wave mode. If the line-of-sigh magnetic field component does not change the sign, the mode conversion is negligible in most cases; thus, we adopt $\eta_{\mathrm{L, R}}(f,  z)=0$ in such cases, so radiation transfer equation simplifies to a more familiar form:

\begin{equation}\label{rt}
\frac{\mathrm{d}I_{\mathrm{L, R}}(f,  z)}{\mathrm{d}z}=j_{\mathrm{L, R}}(f,  z)
-\varkappa_{\mathrm{L, R}}(f,  z)I_{\mathrm{L, R}}(f,  z),
\end{equation}
i.e., the emission components propagate independently until a quasitransverse (QT) magnetic field layer is met.
In these QT layers the mode coupling takes place; its computation requires the magnetic field gradient, which is obtained from a linear interpolation of the magnetic field values between the two voxels in which the line-of-sight magnetic field direction changes; the solution for the QT layer has the form:
\begin{multline}
I_{\mathrm{R}}^{(i)}(f)=I_{\mathrm{R}}^{(i-1)}(f)Q_{\mathrm{T}}(f)+I_{\mathrm{L}}^{(i-1)}(f)[1-Q_{\mathrm{T}}(f)],\quad\\
I_{\mathrm{L}}^{(i)}(f)=I_{\mathrm{L}}^{(i-1)}(f)Q_{\mathrm{T}}(f)+I_{\mathrm{R}}^{(i-1)}(f)[1-Q_{\mathrm{T}}(f)],
\end{multline}
where the indices $i$ and $i-1$ refer to the current and previous voxels, respectively, and the coupling factor $Q_{\mathrm{T}}$ is computed with the exact equations of \citep{Cohen_1960, Zlotnik_1964} for the mode-coupling in a quasi-transverse magnetic field layer. We call this formalism the ``exact coupling''.
The account of the frequency-dependent mode coupling is an essential enhancement of the fast codes compared with earlier implementations described in \citep{Fl_Kuzn_2010, Kuznetsov_etal_2011}, where only the extreme cases of the weak and strong couplings were implemented. \gx\ uses by default  the exact coupling mode, although the weak and strong coupling modes are also available for testing purpose and can be selected in the corresponding drop-down menu.

If the projection of the magnetic field vector on the line-of-sight is positive, then the right- and left- elliptically polarized components correspond to the X and O modes, respectively; otherwise (if the projection is negative), the correspondence is opposite. Note, that even though the emission escaping the data cube is elliptically (not circularly) polarized in a general case, it is implicitly assumed that it then propagates through a coronal plasma with declining values of both electron density and the magnetic field and so the polarization ellipse evolves towards circularity due to the effect of \textit{limiting polarization} \citep[see, e.g.,][]{Zheleznyakov_1997, Fl_Topt_2013_CED}; thus, the emission arriving at the observer is adopted as purely circularly polarized.

Another enhancement of the new implementation of the fast code is its ability to simultaneously solve the radiation transfer equations for many lines of sight (i.e., to simultaneously compute emission from many pixels) using parallel threads provided by modern multi-core processors. By default, the code determines the number of available threads $N_{thr}$ of the processor and uses $N_{thr}-1$ of them to compute emission, reserving one remaining thread for user interaction with the computer while computing emission; however, the user may change this allocation by explicitly typing the desired number of parallel tasks to the corresponding window of the \gx\ interface. These libraries can be used both within \gx\ and independent of it, so long as the simple calling conventions described in the \gx\ documentation are observed.

By default \gx\ computes $I_{\mathrm{L}}$ and $I_{\mathrm{R}}$ intensities and all possible combinations of them (total power, polarization, Stokes $V$) at one hundred logarithmically spaced frequencies between 1 and 100 GHz, but the range and logarithmic separation settings can be adjusted to different user needs. The nonthermal GS emission is often produced only in a fraction of the datacube designated to model the flaring loop(s), while in the remaining volume of the datacube the thermal free-free emission is produced. Importantly, the radiation transfer including mode coupling is considered along the entire line of sight, so the polarization can change well above the flaring loop. Typically, the time needed to compute the emission from all frequencies in all pixels is about one minute.

\subsection{X-ray emission}\label{sec:x-ray}

For chosen electron spectrum and plasma properties, X-ray routines calculate observable flux of X-rays
at 1 AU.  The total flux from each voxel is the combination of thermal and non-thermal bremshtrahlung radiations. The current version of X-ray codes uses a simplified version of soft X-ray emission from plasma with temperature $T$ and electron ($n_e$) and proton ($n_p$) densities  \citep[e.g., Eq 2.3.14 in][]{2004psci.book.....A},
\begin{equation}\label{eq:xsoft}
  I(\epsilon)=8.1\times 10^{-39}\frac{n_e^2V}{\epsilon \sqrt{T}}\exp\left(-\frac{\epsilon}{k_bT}\right)
\end{equation}
where $I(\epsilon)$  is the photon flux spectrum [photons~cm$^{-2}$~keV$^{-1}$~s$^{-1}$], $\epsilon$ is the photon energy in keV, $V$ is the voxel volume. Similarly, the thin-target hard X-rays are calculated as described in \citep[e.g.][as the recent RHESSI reviews]{2011SSRv..159..107H,2011SSRv..159..301K}
\begin{equation}\label{eq:xsoft_2}
  I(\epsilon)=\frac{n_pV}{4\pi R^2}\int_{\epsilon}^{\infty}\sigma(E,\epsilon)F(E)dE
\end{equation}
where $R$ is 1AU distance, $F(E)$ is the electron flux spectrum [photons~cm$^{-2}$~keV$^{-1}$~s$^{-1}$], $\sigma(E,\epsilon)$ is the bremsstrahlung cross-section. The angle-averaged bremsstrahlung cross-section by \citet{1997A&A...326..417H} is used for computations. The choice of cross-section is determined by the compatibility with RHESSI OSPEX software \citep{2002SoPh..210..165S}, where similar angle-averaged approximations are employed. The total X-ray flux spectrum (both Soft and Hard X-ray), logarithmically spaced between 3 and 300 keV, is determined as the sum along the line of sight for each pixel in X-ray maps. The user can see the resulting X-ray spectrum for each pixel of the map.

X-ray emission code is written as a separate module in IDL and allows further additions and improvements.
Currently, X-ray code includes only electron-ion bremsstrahlung and does not account for Compton
scattering or photoelectric absorption of X-rays in the solar atmosphere. The latter will produce
a broad hump on the photon spectrum around 30-50~keV \citep{1978ApJ...219..705B} and can be
accounted for in RHESSI OSPEX software \citep[see][ for the details]{2006A&A...446.1157K}.

\begin{figure}
\begin{center}
\includegraphics[width=0.35\columnwidth,angle=0]{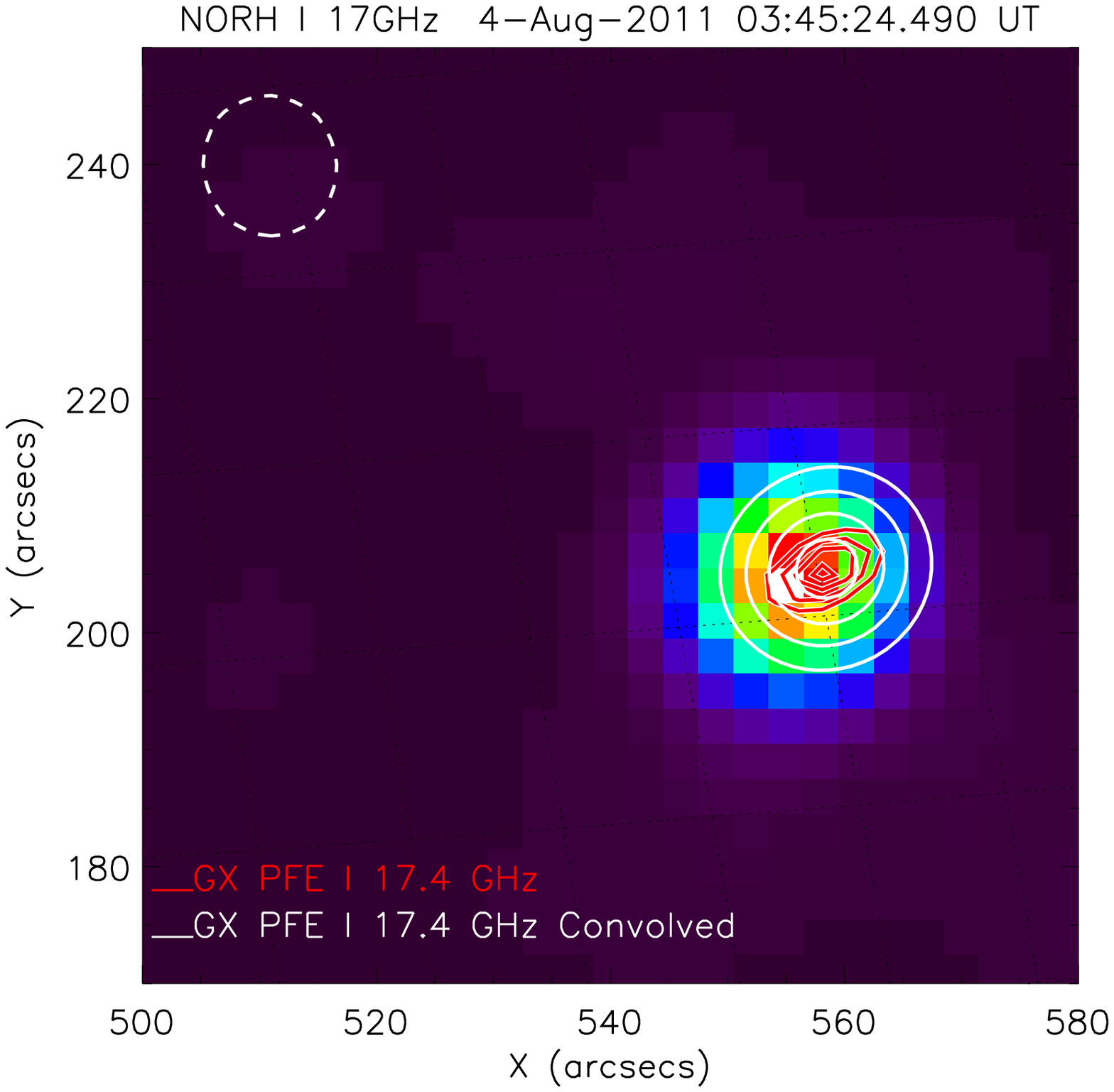}
\includegraphics[width=0.35\columnwidth,angle=0]{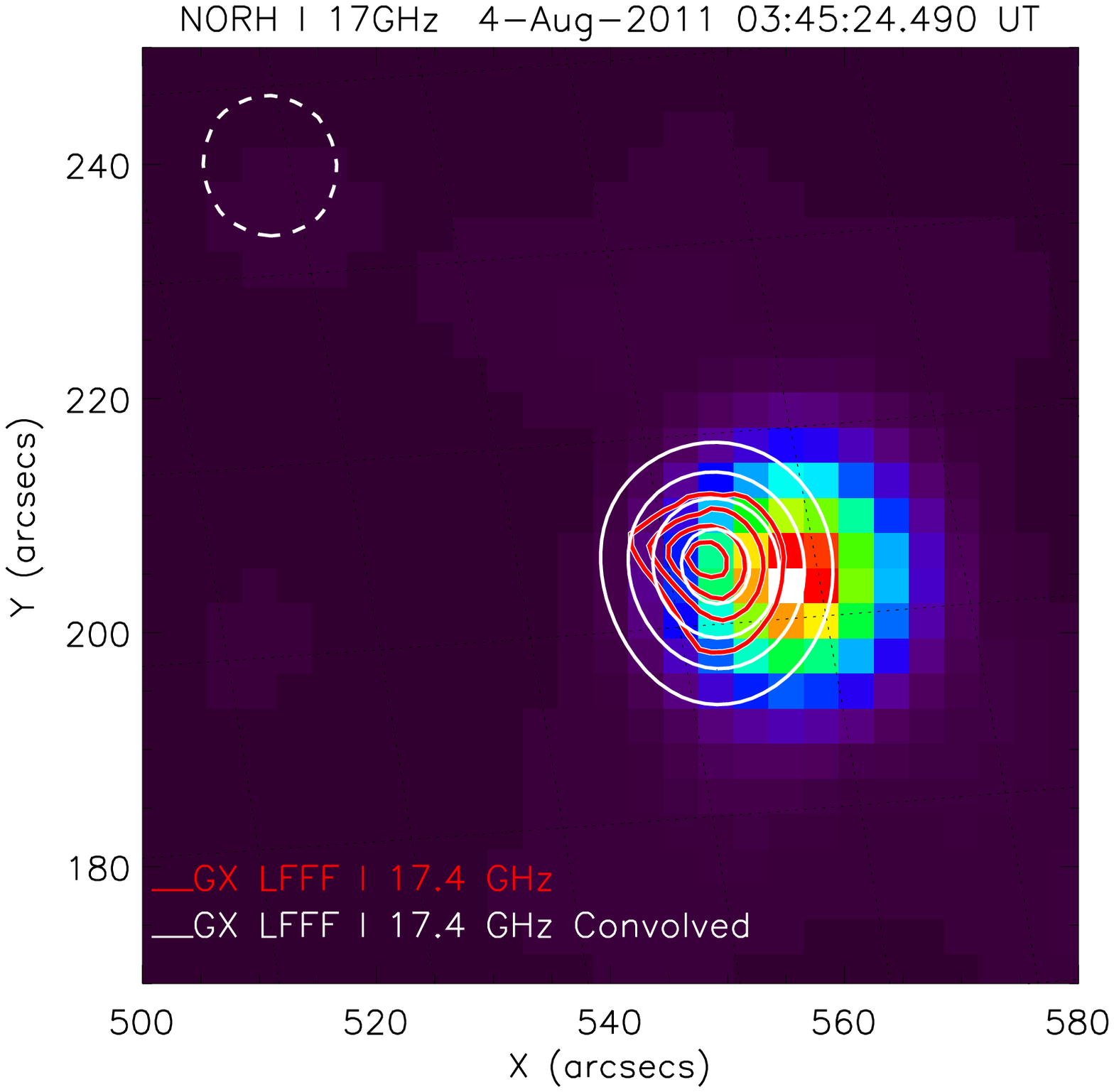}\\
\includegraphics[width=0.35\columnwidth,angle=0]{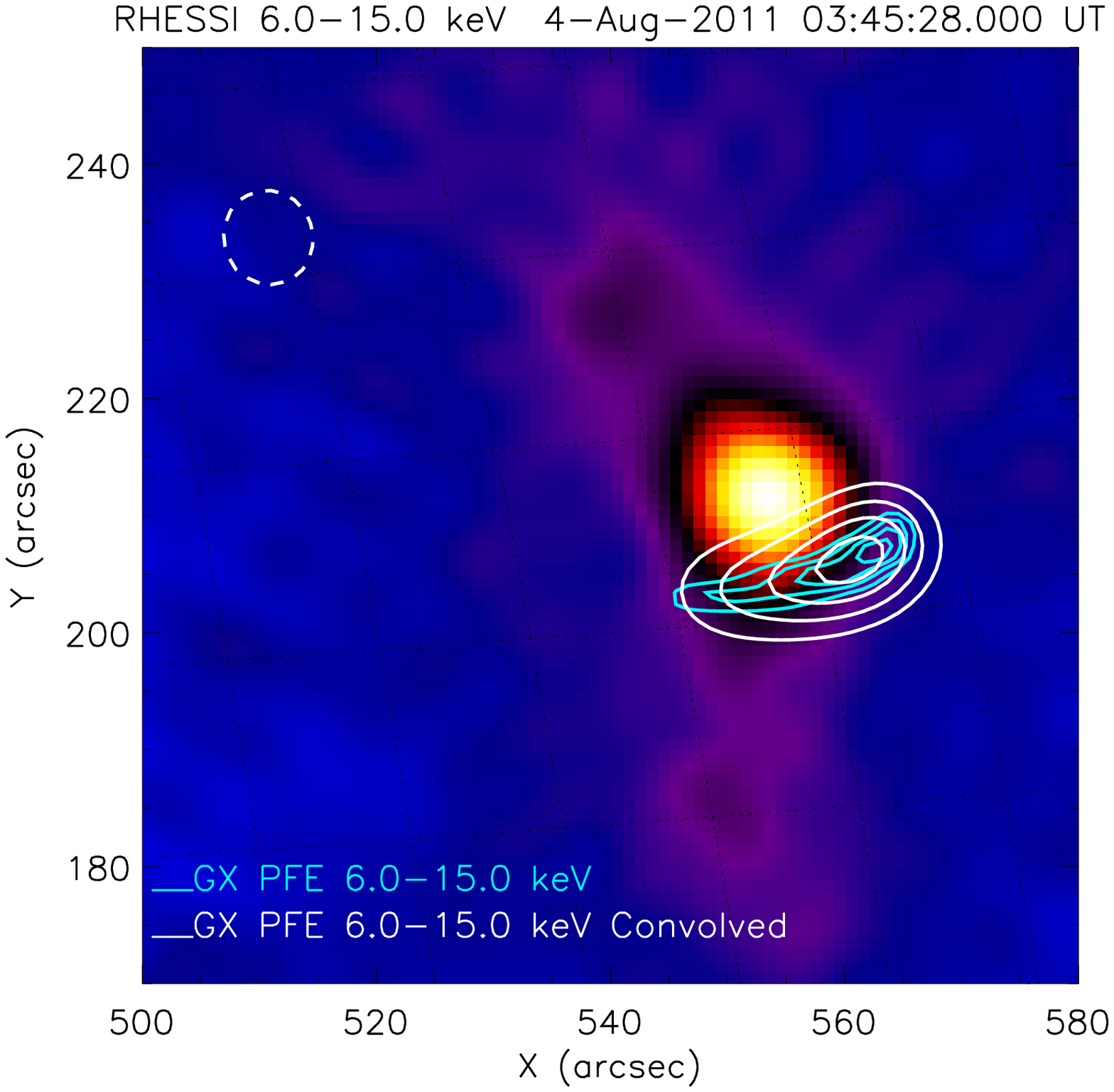}
\includegraphics[width=0.35\columnwidth,angle=0]{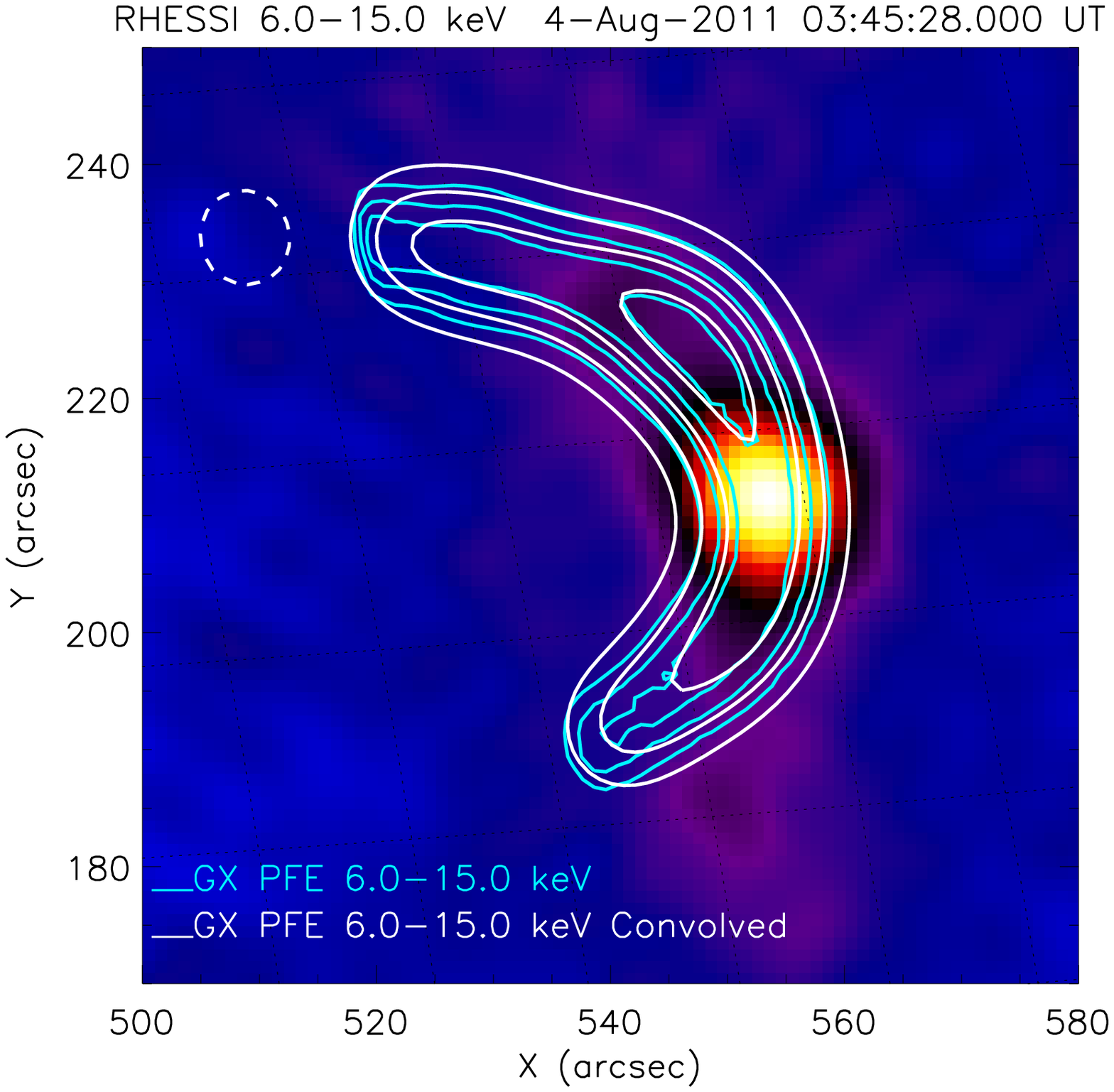}\\
\includegraphics[width=0.35\columnwidth,angle=0]{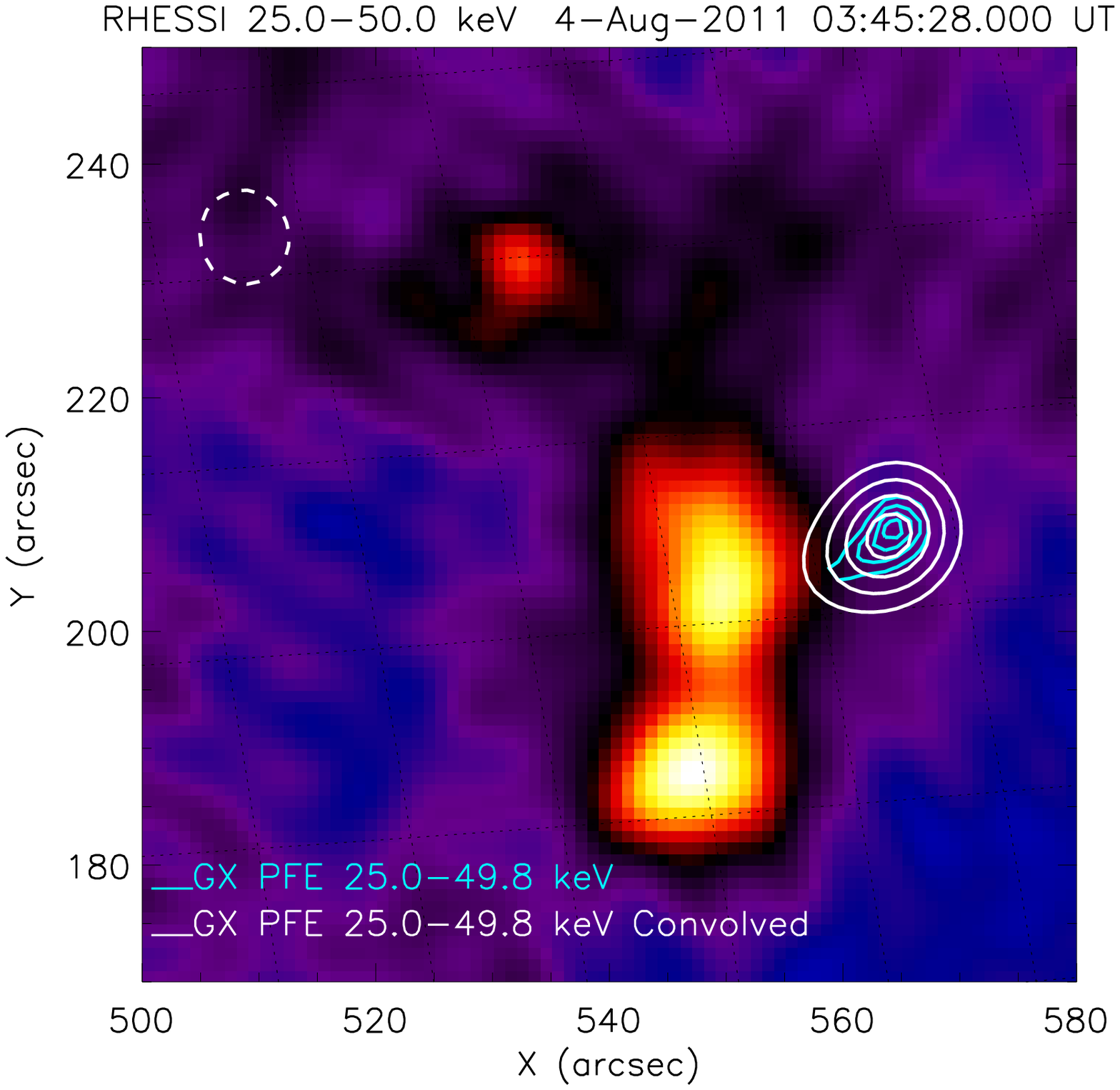}
\includegraphics[width=0.35\columnwidth,angle=0]{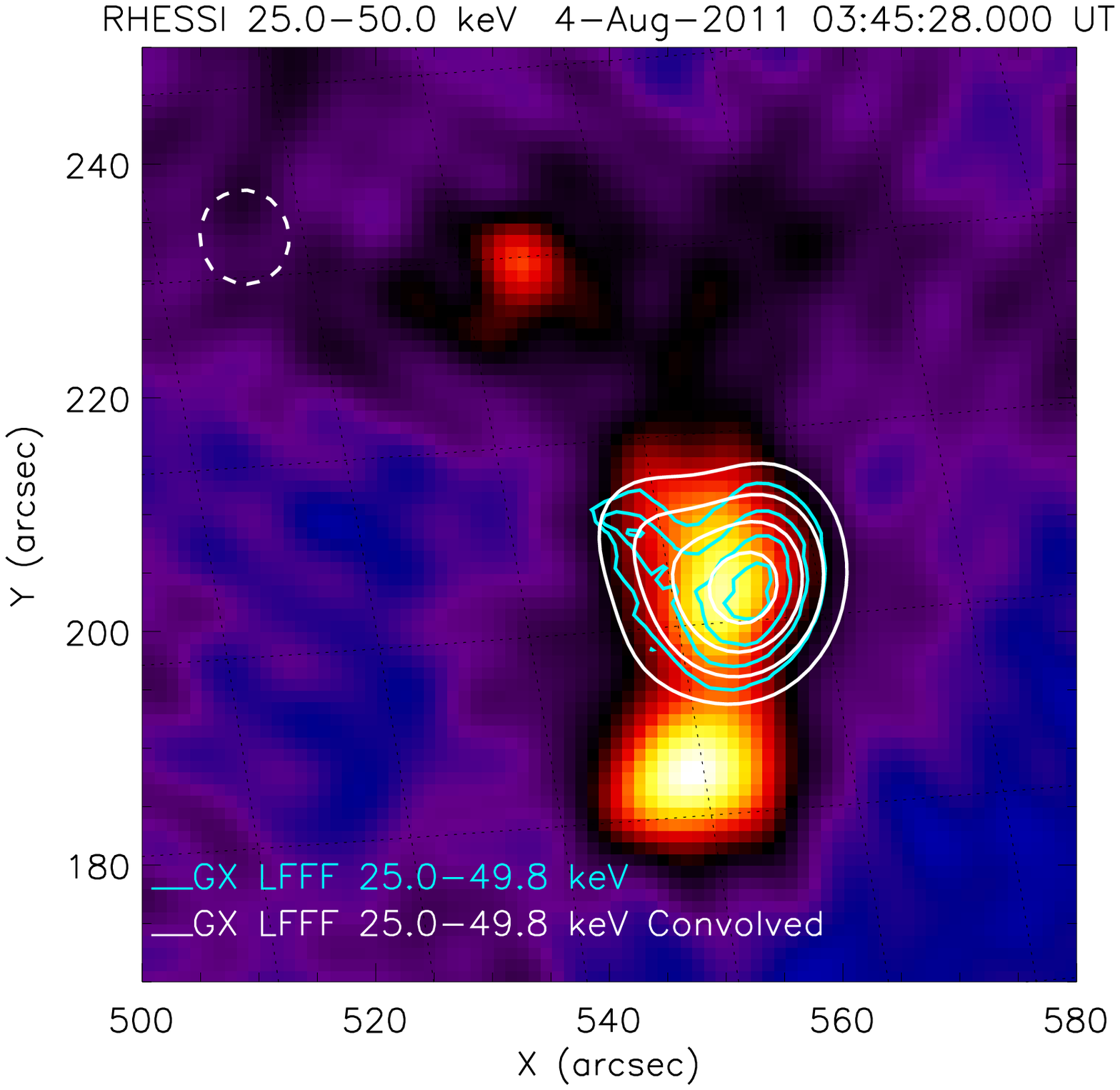}
\end{center}
\caption{\label{gxdatamaps} Top row: \gx\  17 GHz $[30,50,70,90] \%$ synthetic radio emission contours derived from the PFE (left panel) and LFFFE (right panel) magnetic field models displayed against the NoRH 17GHz total intensity maps. Middle row: \gx\  6-15 keV $[30,50,70,90] \%$ synthetic x-ray emission derived from the PFE (left panel) and LFFFE (right panel) magnetic field model displayed against the RHESSI 6-15keV maps.
Bottom row: \gx\  25-50 keV $[30,50,70,90] \%$ synthetic x-ray emission derived from the PFE (left panel) and LFFFE (right panel) magnetic field model displayed against the RHESSI 25-50keV maps. The white contours in each panel were obtained by convolving the \gx output with the NORH and RHESSI instrument circular beams beams, $12"$ and $8"$, respectively.}
\end{figure}

\emph{Example:} In the case of the 4 Aug 2011 solar flare, we have employed the built-in fast gyrosynchrotron codes \citep{Fl_Kuzn_2010} for computing multiple radio maps at a set of microwave frequencies.  The computed 17~GHz $[30,50,70,90] \%$ total intensity contours are displayed in the top row of Figure~\ref{gxdatamaps}, for the PFE (left panel) and LFFFE (right panel) extrapolations, plotted on the corresponding NoRH 17GHz map. Using the built-in thin-target hard X-ray radiation transfer code, we also produced synthetic maps covering the observational range of the RHESSI instrument. The simulated 6-15~keV (middle row) and 25-50~keV (bottom row) $[30,50,70,90] \%$ contours derived from the PFE and LFFFE magnetic field models, are shown in the left and right panels, respectively, against the integrated 6-15 keV and 25-50~keV RHESSI maps. The white radio and X-ray contours in each panel were obtained by convolving the \gx\ output with circular beams, $12''$ and $8''$, respectively.

\section{Model-data comparison}\label{sec:comparison}

As indicated in Figure~\ref{fig:chart}, an essential element of observation-based modeling is the comparison of the model results to the data, and associated iterative adjustment of the model (indicated by dashed lines in Figure~\ref{fig:chart}).  At present this is largely a manual exercise, but \gx\ is designed to aid in the effort by allowing spatial overlays of model and data. This is accomplished by exporting the simulated data cube for a given type of emission to the built-in Plotman display, where Plotman's powerful tools for overlaying images becomes available.

\begin{figure}
\begin{center}
\includegraphics[width=0.48\columnwidth,angle=0]{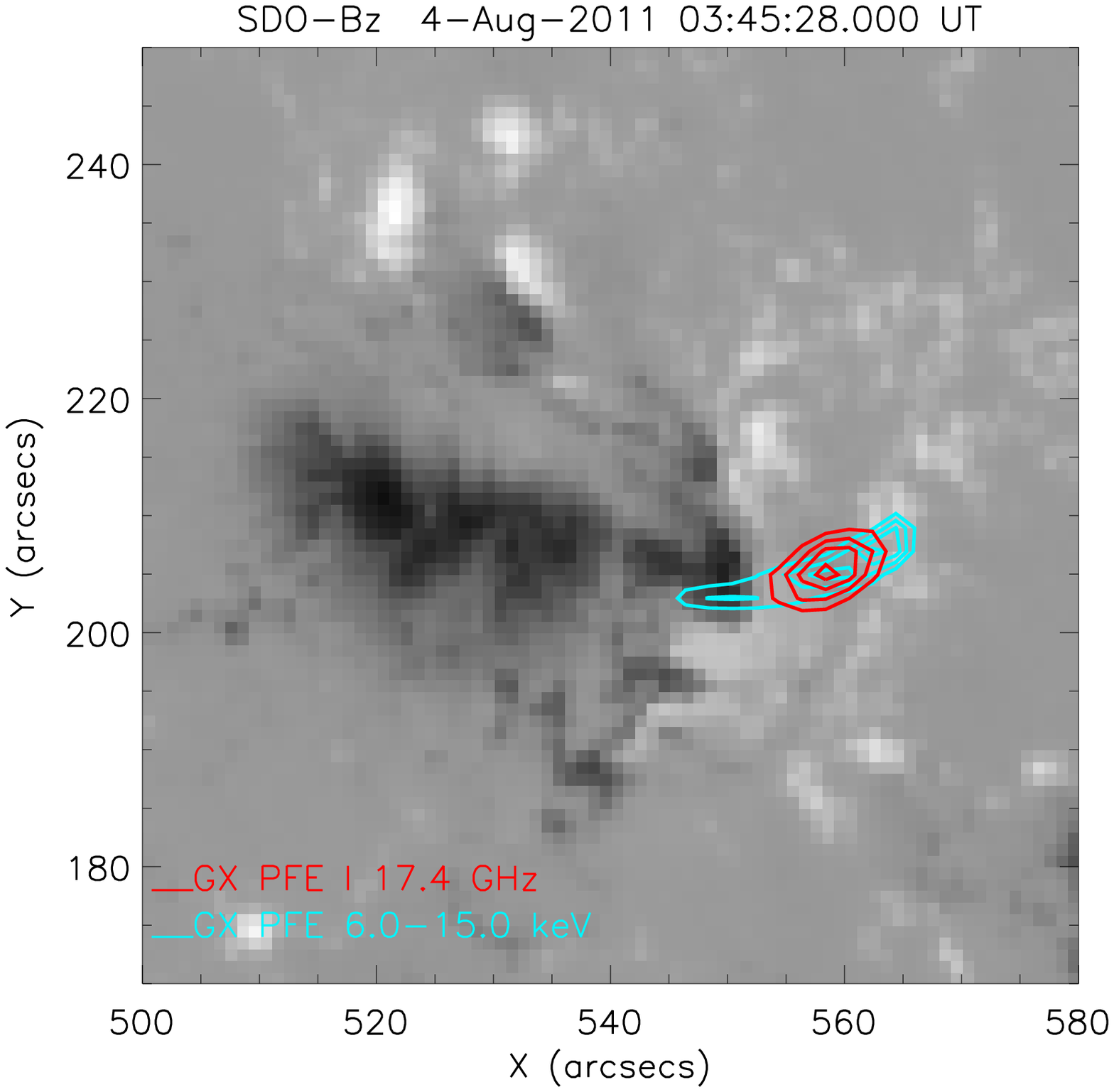}
\includegraphics[width=0.48\columnwidth,angle=0]{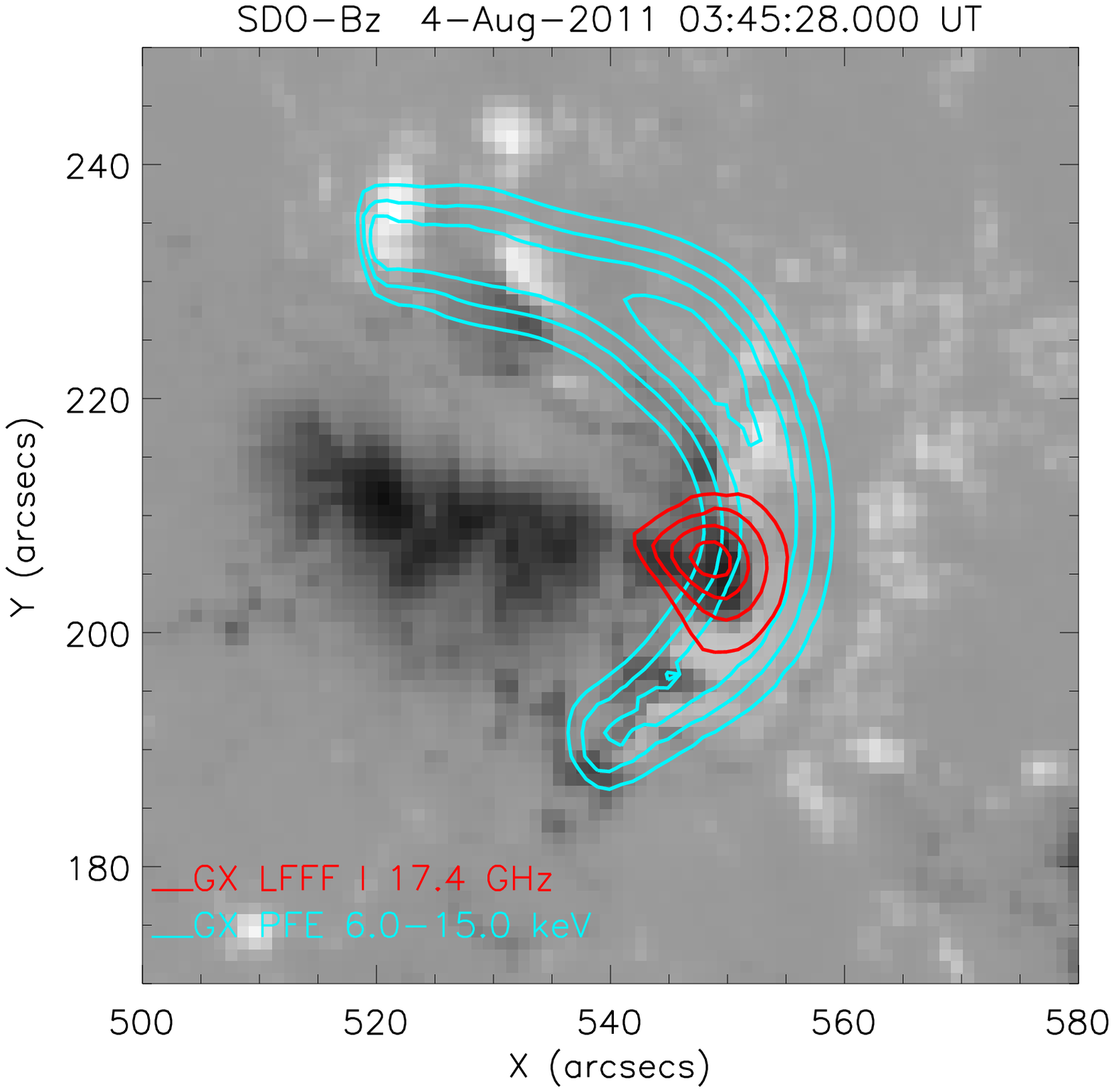}
\\
\includegraphics[width=0.48\columnwidth,angle=0]{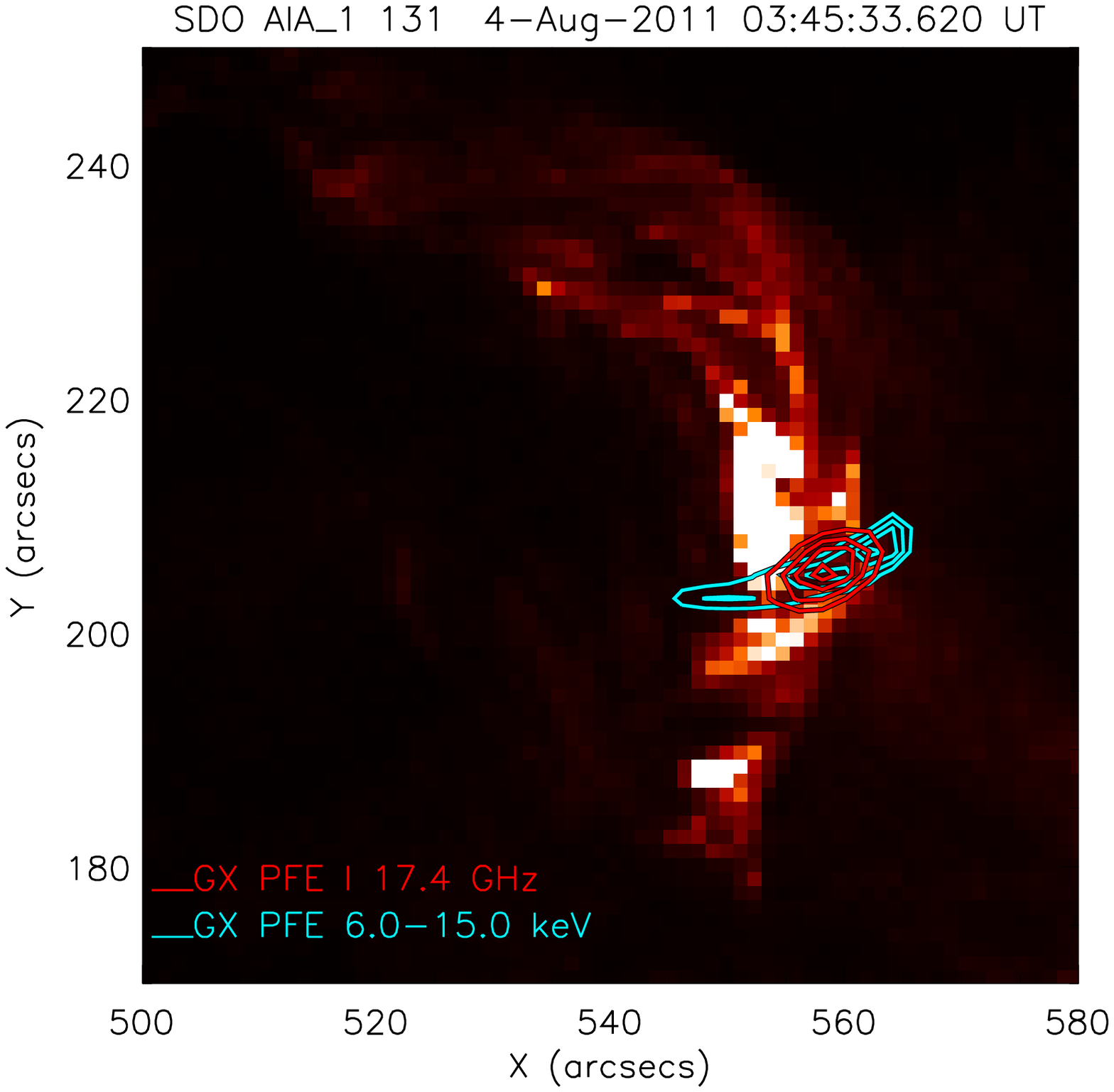}
\includegraphics[width=0.48\columnwidth,angle=0]{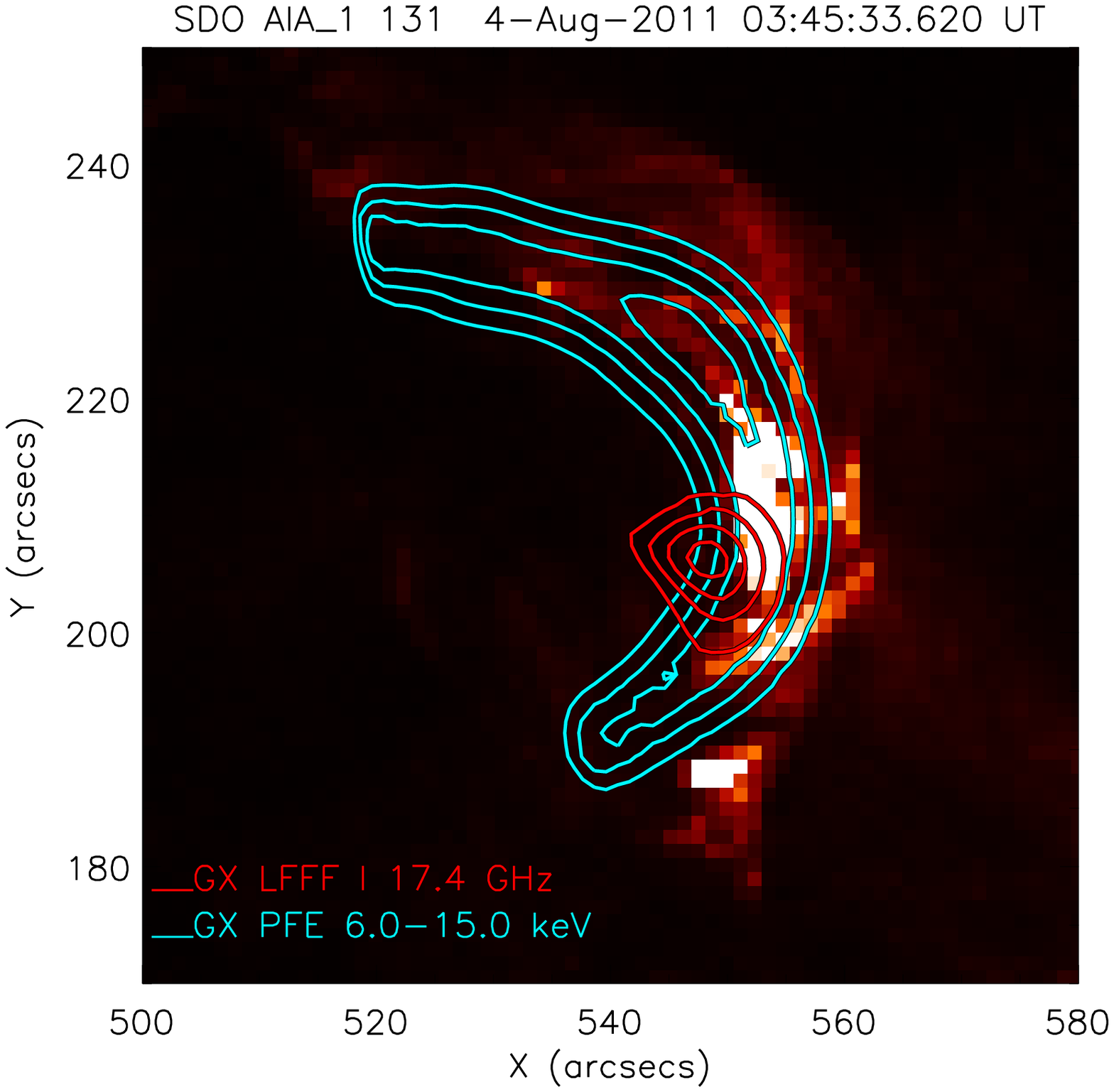}
\end{center}
\caption{\label{gxmaps} \gx\  $[10,30,50,70,90] \%$ simulated emission contours at 17 GHz (solid red) and 6.4 keV (solid blue) derived from the PFE (left panel) and LFFFE (right panel) magnetic field models displayed against the SDO LOS magnetic field (top row) and AIA 131{\AA} (bottom row) maps.}
\end{figure}

\emph{Example:} Model-data comparison is facilitated by \gx, as illustrated in Figure~\ref{gxmaps}, where the simulated radio (red) and X-ray (blue) contours for the PFE and LFFFE models are displayed against the photospheric LOS magnetic field map (top row) and AIA 131 {\AA} image (bottom row). These plots further support the conclusion that the LFFFE model (right column), which follows the neutral line of the active region, produces both radio and X-ray outputs that are in much better agreement with the observational data than the PFE model (left column), which is more confined and transverse to the neutral line. Note that, on close inspection, a subset of thermal EUV loops can be found in the AIA image that are transverse to the neutral line and nearly parallel to the PFE model flux tube, indicating that the magnetic topology is more complex than either the PFE or LFFFE models can capture alone. Nevertheless, the LFFFE model provides the larger source sizes and general orientation that better match the observed nonthermal sources.

\begin{figure*}
\begin{center}
\includegraphics[width=16cm]{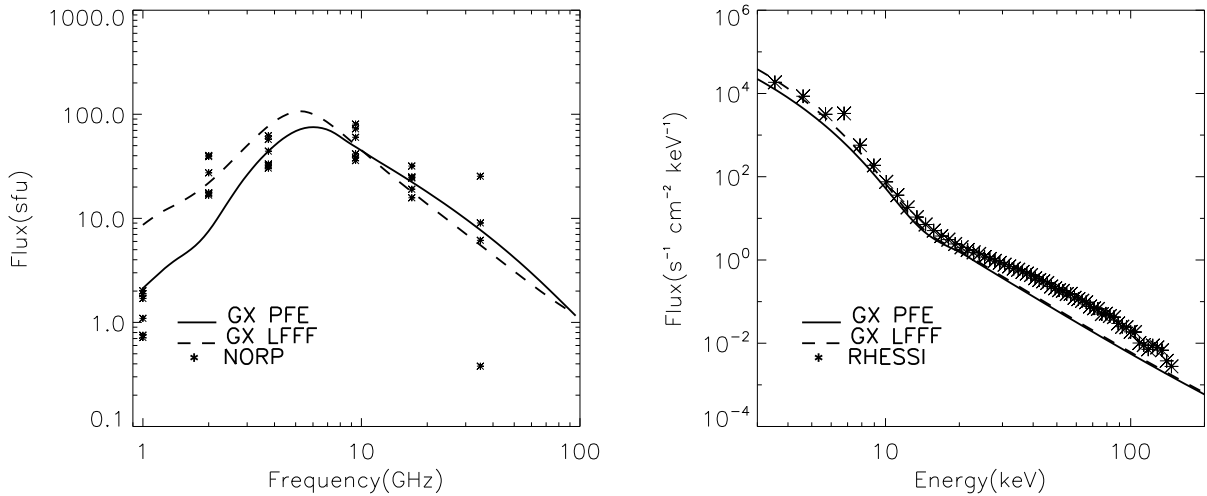}
\end{center}
\caption{\label{data2sim} Left panel: Simulated microwave spectra (lines) for the PFE (solid line) and the LFFFE (dashed line) magnetic field models, versus the observed NoRP spectra (symbols) between 03:45:24UT and 03.45:30UT. Right panel: Simulated X-Ray spectra for the PFE (solid line) and the LFFFE (dashed line) magnetic field models, and the RHESSI spectrum (symbols) integrated between 03:45:28UT and 03:45:49UT. By suitable choice of parameters, the spectra can be made to agree quite well in both cases.}
\end{figure*}

Although spatial overlays are a powerful way to define the magnetic field topology and particle spatial distributions, once these are in rough agreement it is spectral comparisons that show where quantitative adjustments to parameters are needed.  To this end, \gx\ can export to Plotman any spatially resolved spectrum corresponding to a selected spatial location in the simulated image data cube, or the spectrum integrated over the entire field of view. The user can then compare the simulated spectra directly with observed spectra. The result of such analysis is illustrated in Figure \ref{data2sim}, where the FOV integrated synthetic radio (left panel) and X-ray (right panel) spectra derived from the PFE (solid lines) and LFFFE (dashed lines) are compared with the spectra (symbols) observed by NoRH and RHESSI.  On a first iteration, this spectral comparison is generally very poor, and only through adjustment of model parameters on subsequent iterations does this comparison begin to converge.  By suitable adjustment, we were able to arrive at parameters (those given in Table~\ref{distrib_table}) that reasonably reproduce the main features of the observed radio and X-ray spectra for both PFE and LFFFE models, though it may require additional fine tuning of the model parameters to fit a spectrum recorded at multiple time frames.  Once the parameters are adjusted for the spectral fit, one can assess whether the required parameters are physically plausible.

\begin{figure}
\begin{center}
\includegraphics[width=0.42\columnwidth,angle=0]{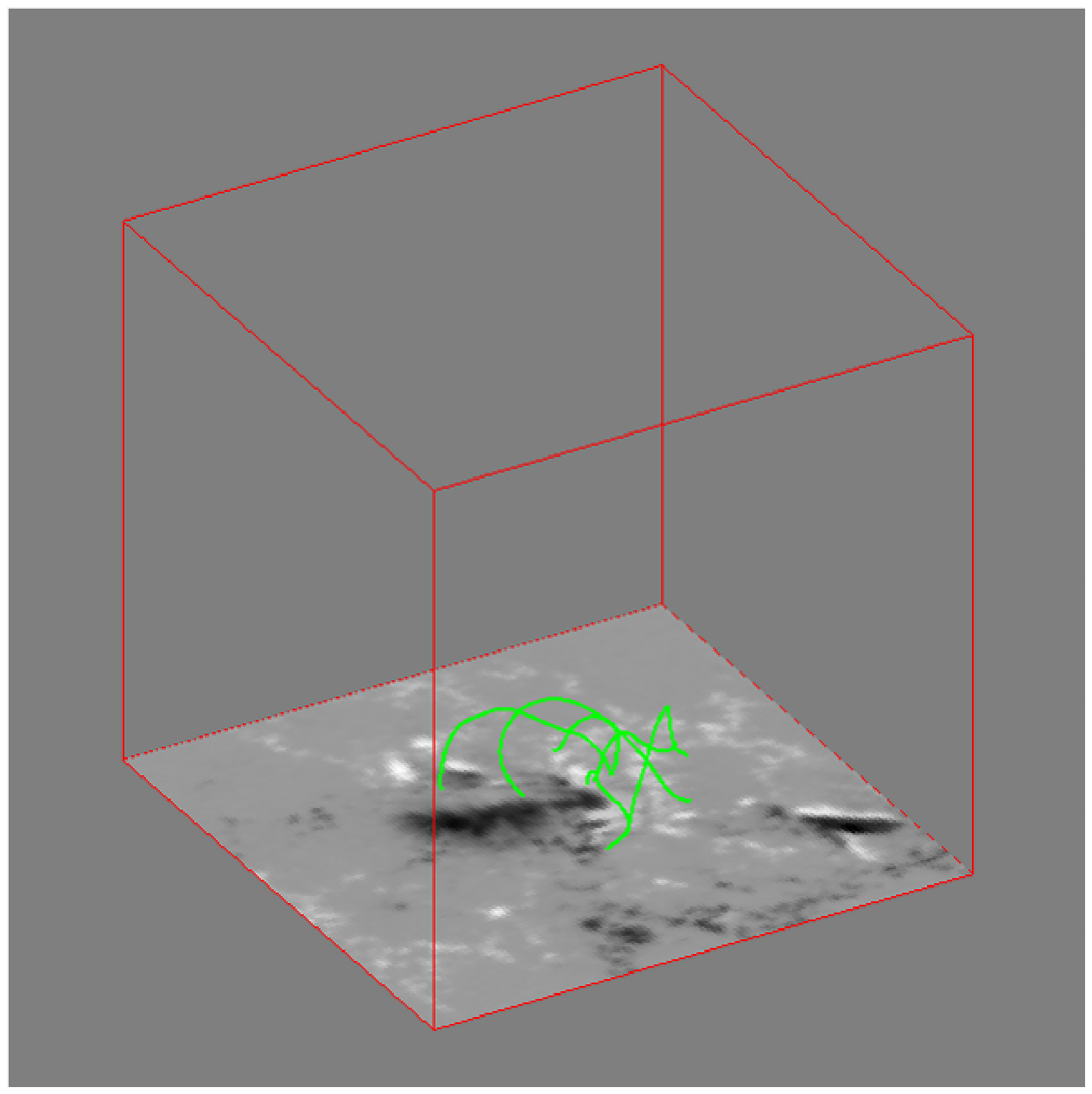}
\includegraphics[width=0.42\columnwidth,angle=0]{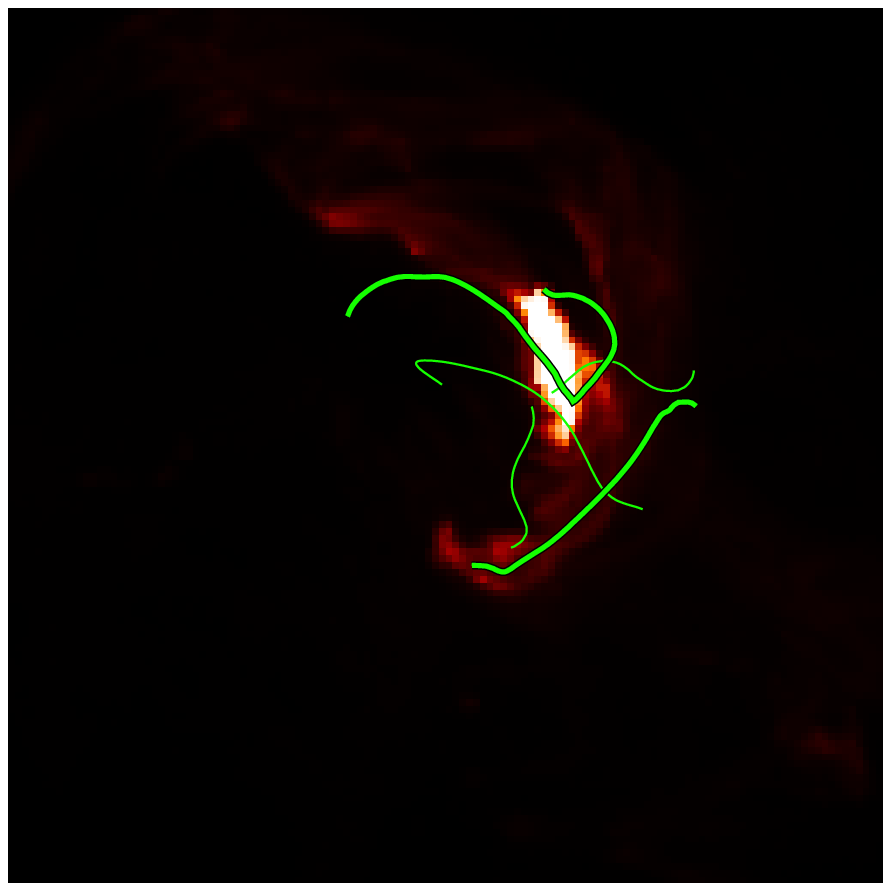}\\
\includegraphics[width=0.42\columnwidth,angle=0]{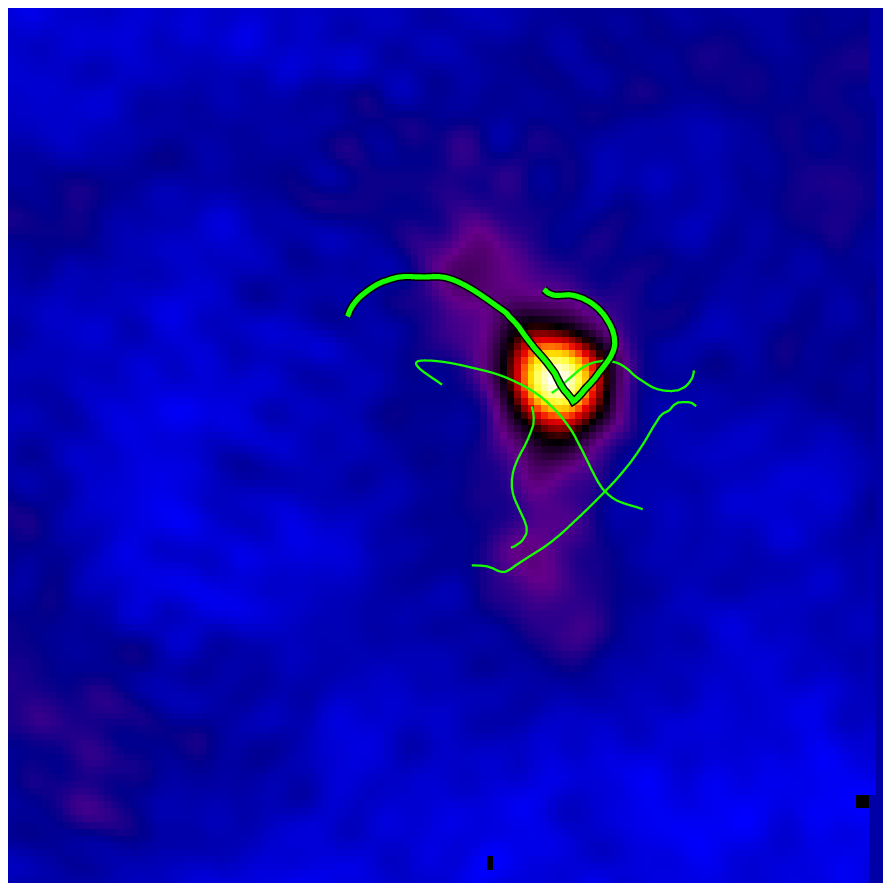}
\includegraphics[width=0.42\columnwidth,angle=0]{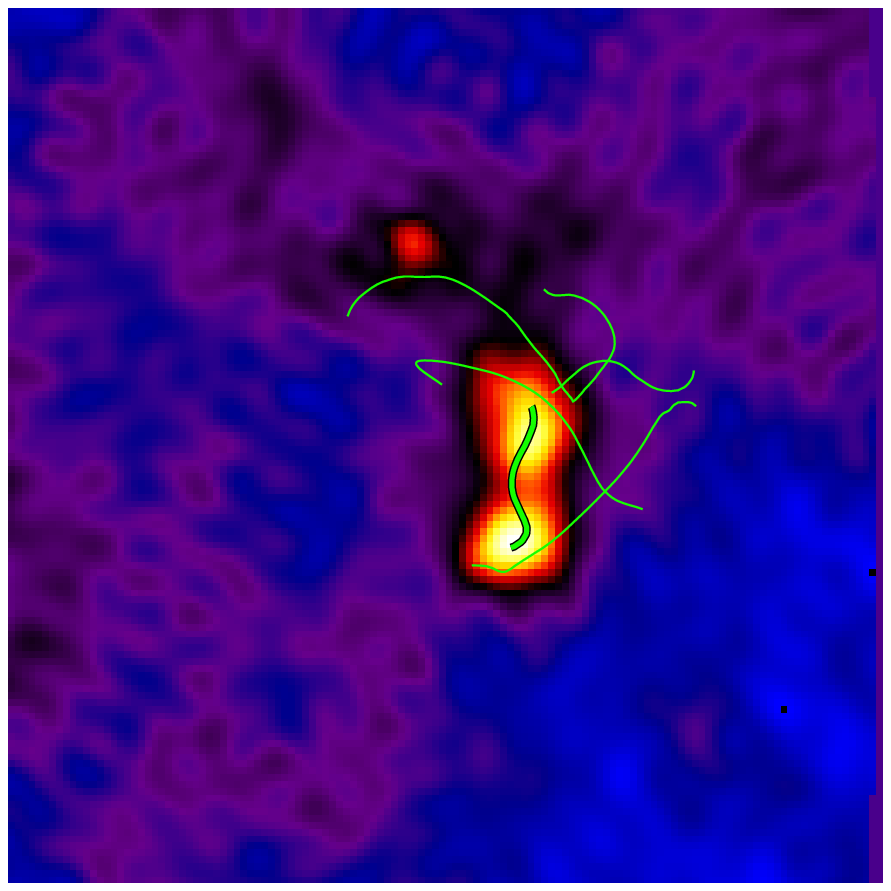}\\
\includegraphics[width=0.42\columnwidth,angle=0]{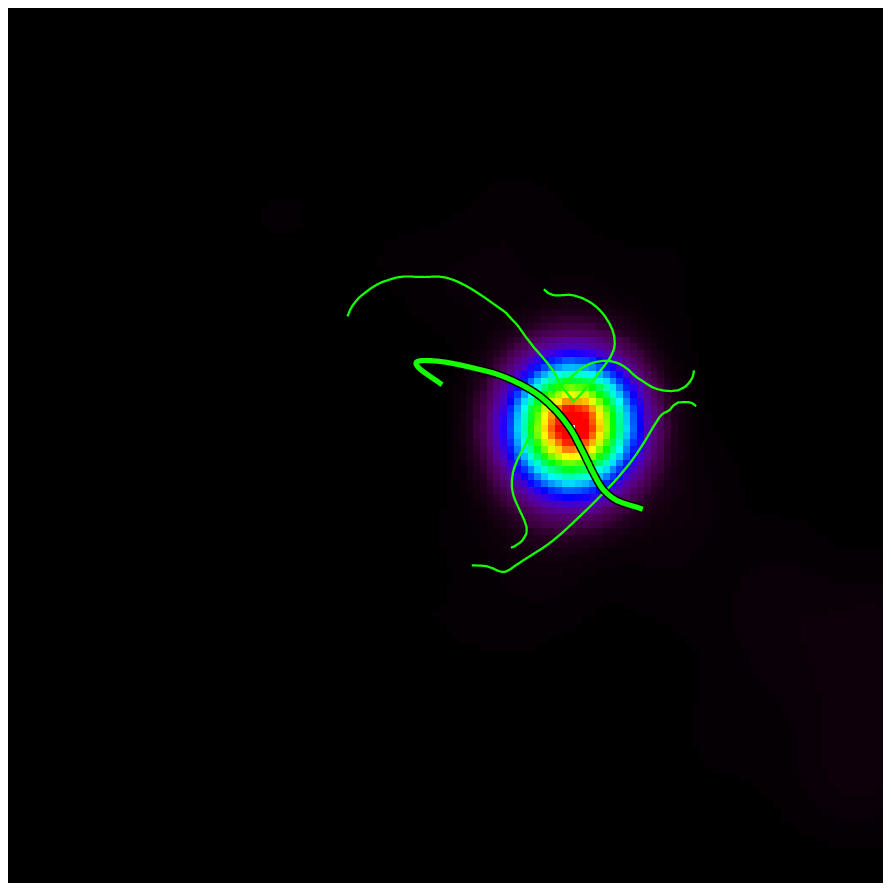}
\includegraphics[width=0.42\columnwidth,angle=0]{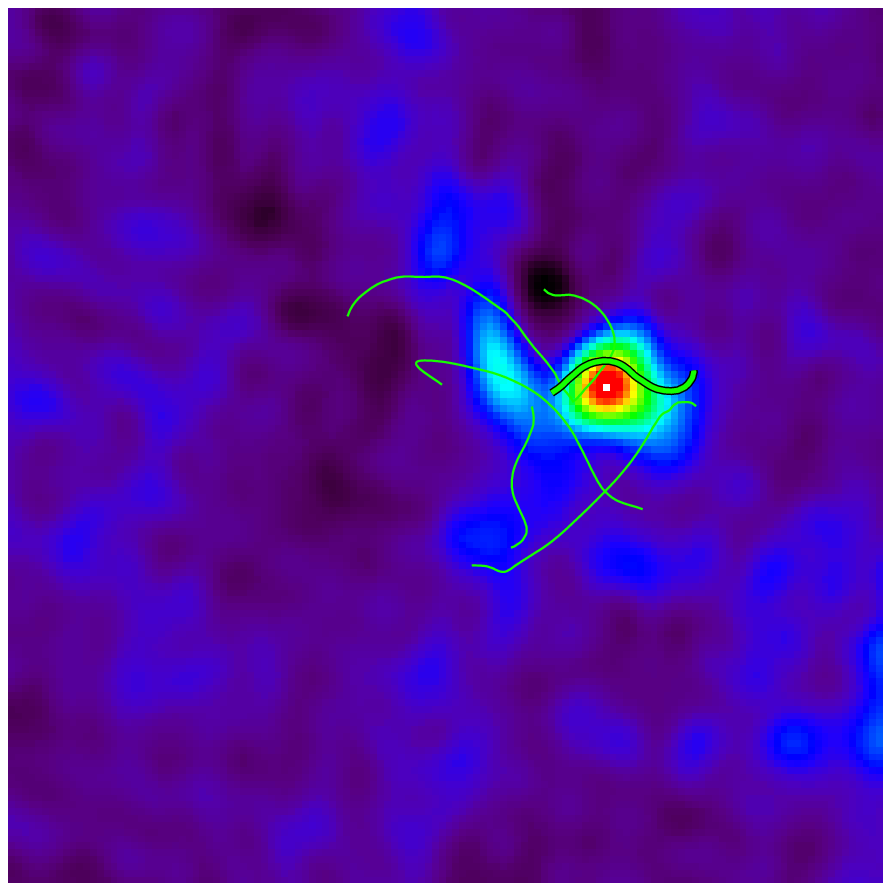}
\end{center}
\caption{\label{nlfff_view} NLFFF field lines. Rows from left to right: 1) HMI and AIA 131{\AA}. 2)RHESSI 6-15keV and 25-50keV. 3) NoRH 17GHz and 34GHz. On each panel, the magnetic field lines that may be associated with the main features in the corresponding reference maps are drawn as thick lines }
\end{figure}

Although we chose what initially appears to be a solar flare with simple geometry, our analysis has revealed a high level of complexity that cannot be fully captured by either PF ($\alpha=0$) or LFFF (constant $\alpha$) extrapolation models.  One might expect a non-linear force-free field (NLFFF) extrapolation model based on vector magnetic field measurements to do a better job.  This also can be handled by the \gx\ tool, and we conclude our 4 Aug 2011 event study by presenting in Figure~\ref{nlfff_view} a set of magnetic field lines provided by an externally performed 3D NLFFF extrapolation data cube provided by Ju Jing (private communication).  The model was imported into \gx\ in the normal way, and a subset of field lines was selected in the vicinity of the flare.
These field lines are shown in a 3D perspective view against the base SDO LOS magnetic field map (top-left),  and in a set of 2D views such as: against AIA 131 {\AA} (top-right), RHESSI 6-15keV (middle-left) and 25-50keV (middle-right), NoRH 17GHz (bottom-left),  and NoRH 34GHz (bottom-right) maps. Unfortunately, the new extrapolation fails to provide a clear single-loop magnetic topology that matches the observations, although it is possible to match the observed radio, EUV, and X-ray structures by considering a few distinct flux tubes. We have not attempted to further analyze this case, which would not fit the purpose and scope of this introductory study, but we point out that \gx\ permits such more-detailed studies that could lead to stringent tests of NLFFFE methods in addition to further insights into the flaring process.

It is quite common to ask how unique the final model is in accounting for all of known observational constraints.  It should be clear from the above that we cannot make any claim to uniqueness, and other combinations of parameters and geometry might equally well fit the observations, but the point of \gx\ is to replace the often-used practice of hand-waving plausibility arguments with quantitatively testable, data-driven modeling.  When more than one solution can fit the observations, one can always seek additional observations, which can provide additional constraints, in order to break the degeneracy and point to a more likely solution. After many iterations on many events, it is expected that use of \gx\ will expose gaps in our physical understanding of flaring events, and eventually lead to a firmer foundation for observationally-driven flare models.

\section{Other functionality and future development of the tool}

Although we mentioned most of the currently available tool functions and capabilities, we demonstrated only a limited subset of them.  Due to the intuitive nature of the interface, we believe the interested user will quickly become familiar with the functions of the tool, as currently available via the SSW distribution. \gx\ will further be enhanced in the future and the upgrades will be added to the SSW. The next anticipated upgrades will have a number of new features including (i) numerically defined thermal models based on a realistic coronal heating model, (ii) ability to compute EUV emission, (iii) ability to compute active-region and quiet Sun  models including gyroresonance and free-free emission from the thermal and quasi-thermal (e.g., $\kappa$-) distributions \citep{Fl_Kuzn_2014}, (iv) a more realistic chromospheric model, (v) the ability to convolve the model images with the point-spread functions of various instruments, and (vi) ability to handle array-defined electron distributions determined from numerically solved particle transport equations, which will also allow to compute the thick-target hard X-ray emission.

\section{Summary}
We have described \gx, and demonstrated its capabilities using a solar flare example. \gx\ provides the framework for data-driven 3D modeling of flaring regions, with the scope of producing synthetic, multi-wavelength microwave and X-ray 2D imaging and spectroscopic data for comparison with observational data.  The extendable, object-based architecture of \gx\ encourages an open-source approach to this common problem in solar physics. The tool is easy to install and use on Windows, Unix and Mac platforms, and it is distributed along with a detailed Help file that contains \emph{Getting Started} guidelines, as well as with a demo data set that may be used to reproduce the results presented in this paper. We have shown how the interactive graphical user interface allows the user to (i) import photospheric magnetic field maps and perform magnetic field extrapolations to almost instantly generate 3D magnetic field models, or alternatively to import field models generated externally, (ii) interactively explore the magnetic topology of these models by creating magnetic field lines and associated magnetic flux tubes, (iii) populate the flux tubes with user-defined nonuniform thermal plasma and anisotropic, nonuniform, nonthermal electron distributions; (iv) produce multi-wavelength images and spectra of radio and  X-ray emission calculated from the model to investigate their spatial and spectral properties, and (v) compare the model-derived images and spectra with observational data. \gx\ integrates shared-object libraries containing fast gyrosynchrotron emission codes developed in FORTRAN and C++, soft and hard X-ray codes developed in IDL, a FORTRAN-based potential-field extrapolation (PFE) routine and an IDL-based linear force free field extrapolation (LFFFE) routine. The extendable, interactive interface allows the addition of any user-defined IDL or externally-callable radiation code that follows our simple interface standards, as well as user-defined magnetic field extrapolation routines.

As an example, we have analyzed the 3D structure of an initial peak of a solar flare observed on 4 Aug 2011, using data from different spacecraft and ground-based observatories augmented by modeling using \gx. This analysis has revealed a high level of complexity in the flaring volume and allowed us to quantify the main flaring loop producing the bulk of the nonthermal microwave and X-ray emission. This flaring magnetic loop is clearly non-potential (in the context of a linear force-free model, it requires a force-free parameter of order $\alpha\sim6.8\times10^{-10}~\mathrm{cm}^{-1}$). The length of the central field line forming the loop is $l\approx6.4\cdot10^9$~cm; the magnetic field being $\sim-310$~G and $\sim520$~G at the footpoints and $\sim150$~G at the looptop; see Figure~\ref{fieldline_view} for greater detail and the loop shape. The thermal plasma in the flaring loop has the density $n_{th} \sim 3\times 10^{10}$~cm$^{-3}$ (which, given the flaring tube volume, yields the emission measure of $EM=4\times10^{48}$~cm$^{-3}$) and $T\sim 20$~MK and a gaussian transverse distribution with the parameters listed in Table~\ref{distrib_table}, see Figure~\ref{fluxtube_lfffe}, left column for greater detail.

The fast electrons are concentrated at or near the looptop as often seen in flares \citep[e.g.,][]{2002ApJ...580L.185M}. Both longitudinal and transverse distributions are well described by gaussian functions with the parameters listed in Table~\ref{distrib_table}, see also Figure~\ref{fluxtube_lfffe}, right column. The number density of the fast electrons is $n_{nth}\simeq10^7$~cm$^{-3}$ between $E_{\min}=20$~keV and $E_{\max}=3$~MeV; the spectral index is $\delta\approx3.1$.
We emphasize that recovering the 3D distributions of the flaring loop physical parameters has been made possible by using the advancements of our powerful modeling tool, \gx, which we have specifically developed for this purpose and made freely available via the SSW distribution\footnote{\url{www.lmsal.com/solarsoft/ ssw\_packages\_info.html}}.

We indicated by dashed lines in Figure~\ref{fig:chart} and mentioned in section~\ref{sec:comparison} that a given model can be brought into better agreement with data through an iterative process of comparison of simulated images with the data, and subsequent adjustment of the model.  A simple example of this has been seen here, where we first try and then reject the PFE magnetic model and turn to the LFFFE with multiple, iterative adjustments to ultimately find a suitable alpha.  Ultimately, one can envision the development of more-sophisticated methods of quantitative comparison and model adjustment, which could lead to improved magnetic field extrapolation algorithms, particle acceleration models and so on.

In addition to the described functionality, the tool has other features (and more are planned to be added soon) including the ability to investigate spatially resolved spectra from a given pixel, which will be especially valuable when the imaging spectroscopy data has become available. Such data are anticipated from instruments like the Jansky Very Large Array (VLA), Expanded Owens Valley Solar Array (EOVSA), upgraded SSRT, CSRH, and eventually FASR.

\acknowledgments
This work is supported by NSF grants AGS-1250374, AGS-1262772, AST-1312802 and NASA grants NNX14AC87G, NNX14AK66G to New Jersey Institute of Technology. E. P. K. was supported by STFC grant. Financial support by the European Commission to E. P. K. through the FP7 HESPE network (FP7-2010-SPACE-263086) and to E.P.K. and A.A.K. through the Marie Curie International Research Staff Exchange Scheme "Radiosun" (PEOPLE-2011-IRSES-295272) is also gratefully acknowledged.
This work also benefited from workshop support from the International Space Science Institute (ISSI). We are grateful to Dr. Pal'shin for his help with the Konus-\textit{Wind} data.
\bibliographystyle{apj}
\bibliography{refs_rhessi,fleishman,GX_bib}

\end{document}